\documentclass[aps,prb,reprint,superscriptaddress,longbibliography,floatfix]{revtex4-2}

\usepackage[utf8]{inputenc}
\usepackage[T1]{fontenc}
\usepackage{amsmath,amssymb,bm}
\usepackage{graphicx}
\usepackage{xcolor}
\usepackage{dcolumn,placeins}
\usepackage[caption=false]{subfig}
\usepackage[colorlinks=true,allcolors=blue]{hyperref}

\newcommand{\figureplaceholder}[1]{%
  \fbox{\begin{minipage}[c][3.1cm][c]{0.92\linewidth}
  \centering\textcolor{red}{#1}
  \end{minipage}}}
\newcommand{\maybegraphics}[2]{%
  \IfFileExists{#1}{\includegraphics[width=#2]{#1}}{%
  \figureplaceholder{Missing figure: \nolinkurl{#1}}}}

\begin{document}

\title{How Much Do We Understand Ce \texorpdfstring{$L_3$}{L3} XANES?
Ab Initio Insights into Configuration-Specific Screening at the
Multiplet--Continuum Frontier}

\author{Alessandro Mirone}\email{mirone@esrf.fr}
\affiliation{European Synchrotron Radiation Facility, 71 Avenue des Martyrs,
F-38000 Grenoble, France}

\begin{abstract}
How can local multiplet signatures, collective core-hole
screening, and an extended photoelectron be reconciled within
a consistent interpretation of Ce $L_3$ x-ray absorption
near-edge spectra?
Motivated by the question raised by Kvashnina
[Chem. Eur. J. \textbf{30}, e202400755 (2024)], we address
this challenge using XSpectruplet, which couples local
multiplet dynamics to continuum photoelectron propagation in
self-consistent Kohn--Sham screening potentials.
For Ce$^{4+}$ in CeO$_2$, our heuristic search identifies
primary and secondary stationary screening states whose
calculated white-line maxima lie about 7.3~eV apart after
including their independently calculated total-energy
difference, yet both retain the same effective local
spectroscopic $4f^0$ sector.
Their squared overlaps with the unrelaxed sudden reference
give approximate statistical weights of 63\% and 30\%,
respectively, within an independent-channel description.
Their contrasting screening responses arise from collective
rehybridization of occupied orbitals, rather than occupation
of an additional localized $4f$ spectator, and produce
contributions in both the main white-line and satellite regions.
Ce$^{3+}$ in monazite CePO$_4$ provides the contrasting case
of a localized $4f^1$ spectator participating explicitly in
the multiplet dynamics.
Independent relaxation of neutral $4f$ and white-line states
gives a CeO$_2$ pre-edge--white-line gap of 10.51~eV, close
to the experimental separation of approximately 10.7~eV.
Its excess over the fixed-background spectral gap defines a net
correction, which we transfer heuristically to CePO$_4$ without
changing its local multiplet structure. This gives an estimated
CePO$_4$ gap of 6.38~eV, compared with approximately 8.1~eV
experimentally.
These results provide a step toward a unified interpretation
of the Ce $L_3$ edge: distinct spectral contributions can
arise from collective screening without a change in the
effective local configuration.
The remaining near-edge structure motivates a broader
exploration of stationary screening states
within the same multiplet--continuum framework.
\end{abstract}

\maketitle

\section{Introduction}
\label{sec:introduction}
How much of the Ce $L_3$ near-edge structure do we understand once local
multiplets, core-hole screening, and an extended photoelectron coexist?

The question raised by Kvashnina~\cite{Kvashnina2024} concerns not only the
assignment of individual spectral features, but also the consistency of the
physical picture obtained by combining different descriptions of Ce $L_3$
x-ray absorption near-edge structure (XANES). At the lanthanide $L_3$
pre-edge, resonant x-ray emission signatures distinguish $f$-electron
configurations and are reproduced by atomic multiplet calculations
\cite{Kvashnina2011}. In CeO$_2$, the pre-edge resolved by
high-energy-resolution fluorescence detection (HERFD) is assigned to
quadrupole $2p\rightarrow4f$ transitions and described by atomic multiplet
calculations with a local $4f^0$ initial configuration
\cite{Kvashnina2011,Kvashnina2024}. The double structure of the main absorption
edge, by contrast, has historically been described in terms of
$4f^0$--$4f^1L$ configuration mixing and screened or unscreened core-excited
states \cite{Bianconi1987,Kvashnina2024}. Here $L$ denotes a ligand hole.

The interpretation of the main edge has also invoked distinct aspects of the
core-excited problem. Bianconi and co-workers related the final-state
structure to configuration interaction between $4f^n$ and $4f^{n+1}L$
components of the ground state~\cite{Bianconi1987}. Soldatov and co-workers
combined configurational final states with a full multiple-scattering
description, emphasizing that the excited-electron wavefunction samples an
extended atomic cluster~\cite{Soldatov1994}. More recently, relativistic
multireference calculations for CeO$_2$ and cerocene reproduced the observed
spectra and associated different relaxed core-excited states with
$4f^0$ and $4f^1$ subconfigurations~\cite{Sergentu2021}. These studies place
different emphasis on ground-state configuration mixing, relaxed
core-excited wavefunctions, and photoelectron propagation.

Configuration labels, formal oxidation states, and orbital-projected
populations describe different aspects of the electronic structure.
Core-level line shapes need not imply mixed valence: CeF$_4$ remains
tetravalent while retaining covalent $4f$ admixture \cite{Kaindl1987}.
Pressure-dependent $L_3$ spectra of CeO$_2$ and CeF$_4$ were likewise
interpreted in terms of core-hole-induced many-body effects associated with
$4f$--ligand covalency, compatible with tetravalency \cite{Kaindl1988}.
In CeO$_2$, O $2p$ mixing with Ce $4f$- and $5d$-derived bands has been
quantified by O $K$-edge spectroscopy and electronic-structure calculations
\cite{Minasian2017}. A complementary distinction concerns the relation
between ground-state and core-excited configurations. Ce(IV)
imidophosphorane complexes exhibit a multipeaked $L_3$ edge while their
calculated ground states remain essentially single-determinant, with the
additional structure attributed to multiconfigurational excited states
\cite{Tateyama2025}. How are these descriptions related, and what does an
electronic configuration mean in each of them?

We address this question by distinguishing the local shell
configuration entering the multiplet problem from the collective
rearrangement of the occupied orbitals that screens the core hole.
In this description, an orbital-projected population is not identified with
the number of localized spectator electrons retained in the explicit
multiplet space.
In the interpretation developed here, the two CeO$_2$ screening
states share the same local spectroscopic $4f^0$ sector despite their
different self-consistent DFT backgrounds: their distinct screening
responses do not require an additional localized $4f$ spectator in
the explicit multiplet space. The photoelectron participates in the
local multiplet dynamics while propagating through the electronic
background established by each screening state.

To implement this description, we use XSpectruplet~\cite{Mirone2026,xspectruplet}
within the projector-augmented-wave (PAW) framework of Quantum ESPRESSO
\cite{QE-2009,QE-2017,Blochl1994,KresseJoubert1999}. The importance of the
many-electron response to sudden core-hole creation for XAS transition
amplitudes has also been demonstrated by determinant-based approaches
\cite{Liang2017,LiangPrendergast2018}. In our construction, the occupied
ground-state subspace is transported into the core-hole PAW representation
without orbital relaxation, defining a common sudden reference. Independently
optimized stationary screening determinants are compared with this reference
through determinant overlaps evaluated in the PAW metric. Each screening
determinant defines a separate potential for the spectral calculation, which
couples a local many-body sector to a plane-wave representation of the
photoelectron.

We locate stationary solutions of the Kohn--Sham equations through
self-consistent field (SCF) iterations. Where needed, the initial
maximum overlap method (IMOM) selects the occupied orbitals according
to their overlap with a fixed initial reference, rather than solely
by energy ordering~\cite{Gilbert2008,Barca2018}. These iterations
constitute a search for stationary states, not a simulation of
time-dependent relaxation.

Once the stationary solutions have been found, determinant overlaps
quantify both their relation to the sudden reference and their mutual
similarity. In an exact treatment, stationary states of the core-ionized
system would be eigenstates of a common Hermitian many-electron
Hamiltonian, acting on all electronic coordinates and retaining the
electron--electron interaction explicitly. States at different energies
would therefore be rigorously orthogonal. In the present mean-field
description, however, the screening response is represented by
independently optimized Kohn--Sham determinants, each associated with
its own self-consistent potential. Their mutual orthogonality is not
imposed and need not follow from these separate optimizations.
For CeO$_2$, the mutual squared overlap between the two screening
determinants is nevertheless small, while both retain appreciable
overlap with the same sudden reference.

To combine their spectral contributions, we adopt an independent-channel
approximation to the sudden excitation: the squared overlaps with
the sudden reference are used as approximate statistical weights
for an incoherent sum of the corresponding spectra. This construction
neglects residual nonorthogonality and coupling between distinct
screening channels, while retaining the coupling between the local
multiplet and the photoelectron within each channel.

We first establish the two CeO$_2$ screening states, characterize their
relation to the common sudden reference, and examine their spectroscopic
consequences. We then resolve the local dipole (E1) and quadrupole (E2)
excitation channels and examine the effect of Brillouin-zone sampling on the
calculated spectra. For the primary CeO$_2$ screening background, we compare
the pre-edge--white-line separation obtained from the fixed spectral
Hamiltonian with the energy difference between independently relaxed neutral
final states. Finally, monazite CePO$_4$ provides an open-shell $4f^1$
counterpart with an explicit localized spectator electron. Detailed
constructions and numerical parameters are collected in the appendices and
Supplemental Material.

\section{Physical framework and method overview}

The main ingredients of the calculation are a common sudden reference,
the optimized screening references, and the coupled spectral states
constructed for each screening background.
The electronic backgrounds are calculated using PBE+$U$ in the
Dudarev formulation, with $U_{\mathrm{eff}}=5$~eV applied to the
$4f$ subspace of every Ce site, including the absorbing
site~\cite{pbe,Dudarev1998}.

For the mean-field description of the core-ionized system, we use
the full $Z+1$ equivalent-core approximation. Here $Z$ denotes the
atomic number of the absorber. In the Quantum ESPRESSO calculation,
the absorber is represented by PAW atomic data generated for nuclear
charge $Z+1$ rather than $Z$, including the ionic potential, projectors,
partial waves, and augmentation terms. This equivalent-core
substitution approximates the increased attraction experienced by
the outer electrons after removal of one core electron. For Ce, the
PAW description generated for $Z=58$ is therefore replaced by one
generated for $Z+1=59$. This substitution defines the mean-field
electronic background, while the Ce $2p$ core hole and its multiplet
interactions are retained explicitly in the local spectral problem.

The screening calculations retain the ground-state number of
explicitly treated electrons, with the photoelectron treated
separately. For the two PAW representations used here, the substitution
increases the ionic valence charge by one. Keeping the electron number
unchanged therefore gives a cell charge of $+1$.

We denote the sudden reference by $|\Phi_{\mathrm{sud}}\rangle$.
This Slater determinant is built by transporting the occupied
ground-state orbitals from the original PAW representation into
the $Z+1$ representation without orbital relaxation.
The screening references are obtained by independent optimizations of
the same charged $Z+1$ problem. We denote their determinant
representations by $|\Phi_j\rangle$. Each optimized reference provides
the electronic background entering a corresponding XSpectruplet
Hamiltonian $H_j$.

For CeO$_2$, the screening references are integer-occupied Slater
determinants. For CePO$_4$, the spectral potential is generated from
the fractionally occupied SCF density, whereas determinant-overlap
diagnostics use an integer-occupation representation of the same
orbitals. This distinction reflects the search procedure: KS--IMOM-guided
preparation was followed by two free SCF branches with Gaussian
occupations and freely updated Hubbard matrices, converging to the
primary state and the alternative secondary candidate discussed in
Sec.~\ref{sec:cepo4}. The final fractional density is retained for
spectroscopy. Since fractional occupations do not define a single
Slater determinant, an integer occupied set is selected only for the
overlap diagnostics, without reoptimizing the orbitals or replacing
the spectral potential. The filling convention and preparation details
are given in Sec.~S2.3 of the Supplemental Material.

For a given screening background $j$, XSpectruplet represents a
spectral state as
\begin{equation}
  |\widehat{\Psi}_{j\lambda}\rangle
  =
  \sum_{\alpha}
  |\alpha\rangle_{\mathrm{loc}}
  \otimes
  |\varphi_{j\lambda\alpha}\rangle_{\mathrm{PW}} .
  \label{eq:coupled-spectral-state}
\end{equation}
Here $\lambda$ labels spectral states for that background,
$|\alpha\rangle_{\mathrm{loc}}$ denotes a local many-body
configuration, and $|\varphi_{j\lambda\alpha}\rangle_{\mathrm{PW}}$
is the associated photoelectron amplitude in the plane-wave PAW
representation. The sum over $\alpha$ retains the coupling
between the local multiplet degrees of freedom and the motion
of the photoelectron, accommodating both localized resonances
and extended states.

The background determinant $|\Phi_j\rangle$ is implicit in this
representation. Within our factorized description, the complete
state associated with screening channel $j$ is written as
\begin{equation}
  |\Psi^{\mathrm{tot}}_{j\lambda}\rangle
  \simeq
  |\widehat{\Psi}_{j\lambda}\rangle
  \otimes
  |\Phi_j\rangle ,
  \label{eq:full-state-factorization}
\end{equation}
The tensor product denotes a schematic separation between the active
problem and the frozen DFT background. Fermionic statistics and
channel-dependent exclusions are enforced within the explicit local
sector. The spectral propagation does not explicitly project the
photoelectron onto the complement of the occupied KS background;
its occupied-state character is examined separately in
Appendix~\ref{app:spectral-analysis}.
The same determinant $|\Phi_j\rangle$ accompanies all local
components $\alpha$ and spectral states $\lambda$ within channel $j$.
It remains frozen during the spectral calculation, while its
screening response enters the active problem through the
Hamiltonian $H_j$.

The excitation source is denoted by $|b_\mu\rangle=D_\mu|g\rangle$,
where $|g\rangle$ is the initial local many-body state and $D_\mu$
is the transition operator for the chosen multipole channel and
polarization. For CePO$_4$, the initial local $4f^1$ state $|g\rangle$
is selected by the bond-dependent empirical crystal field described in
Sec.~S8 of the Supplemental Material, while the self-consistent electronic
density defines the screening background.
E1+E2 denotes the coherent response to the combined
excitation operator within a screening channel, rather than the arithmetic
sum of separately calculated E1 and E2 responses. This combination is
distinct from the incoherent sum over different screening references
introduced below. The source construction in the PAW representation is
described in Appendix~\ref{app:paw-source}.
For screening background $j$, the spectral response is
\begin{align}
 I_{j\mu}(\epsilon)
 &= \langle b_\mu|\delta(\epsilon-H_j)|b_\mu\rangle
 \nonumber\\
 &= \sum_\lambda
 |\langle\widehat{\Psi}_{j\lambda}|b_\mu\rangle|^2
 \delta(\epsilon-\epsilon_{j\lambda}),
 \label{eq:spectral-representation}
\end{align}
where $\epsilon_{j\lambda}$ are the eigenvalues of $H_j$ on its
internal XSpectruplet energy scale. Lanczos and Chebyshev recurrences evaluate
this response through repeated applications of $H_j$, without
assembling the full local--continuum matrix.

The relation between screening state $j$ and the sudden reference
is quantified by their squared determinant overlap.
Let $C_{\mathrm{sud},\sigma}$ and $C_{j\sigma}$ contain the occupied
orbitals of the two determinants in the common $Z+1$ representation,
with $\sigma$ labeling spin. The orbitals within each determinant
are orthonormal in the PAW metric $S_{Z+1}$. Their occupied-orbital
overlap matrix is
\begin{equation}
 M_{j\sigma}
 = C_{j\sigma}^\dagger S_{Z+1}C_{\mathrm{sud},\sigma}.
 \label{eq:occupied-orbital-overlap}
\end{equation}
Its singular-value decomposition,
\begin{equation}
 M_{j\sigma}=U_{j\sigma}\Sigma_{j\sigma}V_{j\sigma}^{\dagger},
 \label{eq:occupied-svd}
\end{equation}
has unitary matrices $U_{j\sigma}$ and $V_{j\sigma}$ and nonnegative
diagonal entries $s_{ji\sigma}$ in $\Sigma_{j\sigma}$. The columns
$\bm{u}_{ji\sigma}$ of $U_{j\sigma}$ and $\bm{v}_{ji\sigma}$ of
$V_{j\sigma}$ are the left and right singular vectors associated with
the same $s_{ji\sigma}$. They define corresponding occupied orbital
pairs~\cite{Plasser2016},
\begin{align}
 |\chi_{ji\sigma}\rangle &= C_{j\sigma}\bm{u}_{ji\sigma},\nonumber\\
 |\xi_{ji\sigma}\rangle &= C_{\mathrm{sud},\sigma}\bm{v}_{ji\sigma},
 \label{eq:corresponding-orbitals}
\end{align}
one in the screening state and one in the sudden reference. Each set
remains orthonormal in the PAW metric, and their cross overlaps become
\begin{equation}
 \langle\chi_{ji\sigma}|S_{Z+1}|\xi_{jl\sigma}\rangle
 =s_{ji\sigma}\delta_{il}.
 \label{eq:paired-orbital-overlap}
\end{equation}
Thus each singular value is the overlap of its corresponding pair
and equals $\cos\theta_{ji\sigma}$, where $\theta_{ji\sigma}$ is a
principal angle between the occupied subspaces~\cite{BjorckGolub1973}.
The pairs are linear combinations within the two occupied spaces,
rather than a matching of the original KS eigenstates; rotations within
degenerate singular-value subspaces remain free.

The determinant overlap is therefore
\begin{align}
 P_j
 &= \left|\langle\Phi_j|\Phi_{\mathrm{sud}}\rangle\right|^2
 \nonumber\\
 &= \left|\prod_\sigma\det M_{j\sigma}\right|^2
  = \prod_{i\sigma}s_{ji\sigma}^2,
 \label{eq:determinant-overlap}
\end{align}
Taking its logarithm gives an additive contribution from each pair,
\begin{equation}
 -\log P_j=\sum_{i\sigma}w_{ji\sigma},\qquad
 w_{ji\sigma}=-\log(s_{ji\sigma}^2).
 \label{eq:overlap-loss}
\end{equation}
These contributions show whether the reduction of determinant overlap
is concentrated in a few principal directions or spread over many.
They are invariant under unitary changes of the original occupied-orbital
bases. The orbital-overlap criterion used by IMOM
to select occupied states during SCF is distinct from the
many-electron comparison in Eq.~\eqref{eq:determinant-overlap}.
For a periodic calculation these comparisons can be made separately
within each $\bm k$ block. For CePO$_4$ we report the geometric mean
of the per-$\bm k$ squared overlaps, with the integer occupation
convention defined in the Supplemental Material. This mesh diagnostic
is distinct from the $P_j$ used below for the CeO$_2$ screening channels.

Let $E_{\mathrm{ref},j}$ denote the QE total cell energy of the charged
screening determinant $|\Phi_j\rangle$. To compare and combine
spectra from different screening backgrounds, we include these
reference-state energies when placing the spectra on a common axis.
Taking screening state $1$ as the reference, we define
$\Delta_j=E_{\mathrm{ref},j}-E_{\mathrm{ref},1}$.
For the two CeO$_2$ screening channels, the internal spectral energies
retain the common QE potential convention, using identical PAW
pseudopotentials, cell geometry and charge, without separate Fermi-level
shifts. Both channels also use the same local multiplet Hamiltonian,
with identical shell-average subtractions as specified in Sec.~S8 of
the Supplemental Material; these subtractions do not independently
recenter the screening Hamiltonians.
A feature at internal energy $\epsilon$ in channel $j$ therefore
appears at $E=\epsilon+\Delta_j$, up to an overall energy origin
common to all channels.

If $\epsilon_{\mathrm{WL},j}$ denotes the white-line maximum
on the internal energy scale of channel $j$, the separation
between the white lines of two screening states is
\begin{equation}
 \Delta E_{\mathrm{WL}}
 =
 \Delta E_{\mathrm{ref}}
 +
 \Delta\epsilon_{\mathrm{WL}},
 \label{eq:white-line-accounting}
\end{equation}
where
$\Delta E_{\mathrm{ref}}=E_{\mathrm{ref},2}-E_{\mathrm{ref},1}$
and
$\Delta\epsilon_{\mathrm{WL}}
=\epsilon_{\mathrm{WL},2}-\epsilon_{\mathrm{WL},1}$.
The relative placement of the two white lines thus includes both
the energy difference between the charged references and the
difference between their internal spectral maxima.

Within the sudden, independent-channel approximation, we use
$P_j$ as an approximate statistical weight for screening channel $j$
and construct the combined response as
\begin{equation}
 I_\mu(E)
 \simeq
 \sum_j P_j\,I_{j\mu}(E-\Delta_j).
 \label{eq:screening-weighted-spectrum}
\end{equation}
This incoherent sum neglects residual nonorthogonality and coupling
between different screening channels, while retaining the
local-multiplet--photoelectron coupling within each channel.
The channel spectra retain their calculated source norms;
any overall normalization is applied after combining them.

\section{Results and discussion}
\label{sec:results}

\subsection{Two collective screening states within the same
\texorpdfstring{$4f^0$}{4f0} sector}
\label{sec:screening-states}

\begin{figure*}[t]
 \centering
 \maybegraphics{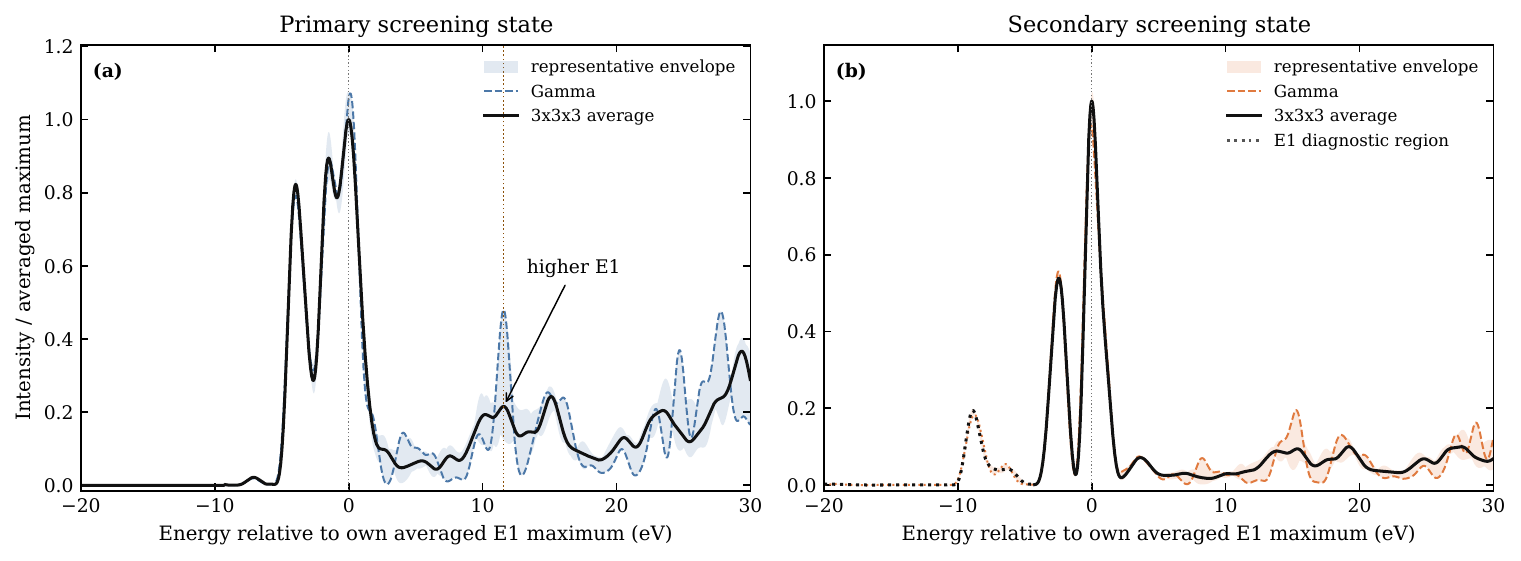}{0.95\textwidth}
 \caption{CeO$_2$ E1 spectra for the primary (a) and secondary (b) screening
 potentials. The dashed curve is Gamma, the colored band spans the eight
 symmetry representatives of a $3\times3\times3$ mesh, and the black curve
 is their multiplicity- and source-norm-weighted average. Each panel is
 aligned and normalized to its own averaged white line; representative
 spectra share that scale. In (b), dotted segments cover the two E1
 features whose dominant diagnostic components are nearly fully occupied,
 and extend to the left edge. The change of style occurs at the minimum
 on their high-energy side. The dotted continuation is a display convention,
 not a pointwise Pauli classification. The primary curves retain the lower display
 cutoff preceding their quadrupolar $4f$ resonance.}
 \label{fig:kdependence}
\end{figure*}

Figure~\ref{fig:kdependence} compares the CeO$_2$ dipole spectra for two
screening backgrounds: the \emph{primary screening state} on the left and
the \emph{secondary screening state} on the right. Each panel shows the
spectrum at the Gamma k-point, the spread across the sampled Brillouin zone, and the
weighted average. Both backgrounds use the same local spectroscopic $4f^0$
sector; the differences arise from the reorganization of the occupied
electronic background that screens the core hole.

The transported CeO$_2$ ground-state determinant supplies the sudden
reference in the $Z+1$ PAW representation. The primary screening state
$|\Phi_1\rangle$ is obtained independently from a charged $Z+1$
Aufbau calculation. Starting instead from the transported determinant and retaining it as
the fixed state-selection reference, a separate KS--IMOM search for the
same charged system identifies the secondary screening state
$|\Phi_2\rangle$.

The Aufbau solution follows ordinary energy ordering, whereas the search
for an alternative stationary screening solution through KS--IMOM is
heuristic.

The PAW-metric squared determinant overlaps with the sudden reference
are $P_1\simeq0.63$ and $P_2\simeq0.30$, while the mutual squared
overlap is
\begin{equation}
  \left|\langle\Phi_1|\Phi_2\rangle\right|^2 \simeq 0.073.
\end{equation}
Both screening determinants thus retain appreciable overlap with
the same physical sudden reference, while their mutual overlap is much
smaller. These results support retaining both contributions in
Eq.~\eqref{eq:screening-weighted-spectrum}, with their residual
nonorthogonality neglected within the independent-channel
approximation introduced above.

The PAW transport procedure is detailed in
Appendix~\ref{app:paw-source}; state-selection procedures and
numerical diagnostics are given in
Appendix~\ref{app:state-selection} and the Supplemental Material.

Using the corresponding orbital pairs defined in
Eq.~\eqref{eq:corresponding-orbitals}, we rank the contributions
$w_{ji\sigma}$ to $-\log P_j$ in Eq.~\eqref{eq:overlap-loss} across both
spins. For the primary state, two principal directions account for 54.76\% of
$-\log P$ and five for 88.27\%; corresponding secondary fractions are
29.87\% and 44.27\%. The primary overlap loss is concentrated in a few
occupied-subspace directions, while the secondary loss is broader. This
gauge-invariant statement is distinct from arbitrary orbital matching and
from a partition of the screening charge.
A principal direction may itself combine many original occupied orbitals;
singular values therefore neither prove nor disprove how that response is
distributed among atomic regions.

The orbital-projected populations reveal how the two screening
states differ. Relative to the primary, the secondary loses
1.67 units of absorber $4f$ weight, gains 1.09 in absorber $5d$,
and gains 0.49 in total O $2p$. The mean projected $4f$ population
on the nonabsorbing Ce sites changes by only 0.001.
The secondary is therefore less screened through absorber $4f$
character, but not devoid of screening.

The final projected populations alone do not establish whether the
additional absorber $4f$ weight forms a localized spectator. We therefore
compare the spin-resolved local $4f$ density matrices of each screening
state with that of the transported sudden reference, using the same
L\"owdin-orthonormalized PAW atomic orbitals throughout. In the primary state,
the total projected absorber population increases from
$n_{4f}^{\mathrm{sud}}\simeq0.65$ to $n_{4f}^{(1)}\simeq1.74$.
Nevertheless, the largest eigenvalues of its local $4f$ density matrices
are only 0.394 and 0.447 in the two spin channels. Since these eigenvalues
give the maximum $4f$ weight obtainable by any normalized combination
of occupied orbitals, the occupied subspace contains no nearly filled
local $4f$ orbital. The corresponding-orbital analysis shows that the
increase is carried mainly by several covalent orbital pairs: seven pairs account for 92.3\%
of the net $4f$ increase, and the two largest contributions raise their
$4f$ weights from approximately 0.11 to 0.37 and 0.42 while retaining
substantial O $2p$ character.

The secondary state exhibits the complementary redistribution. Its
projected absorber population decreases to $n_{4f}^{(2)}\simeq0.067$,
with a largest local $4f$ eigenvalue of only 0.017 in either spin
channel, while its absorber $5d$ population increases relative to the
sudden reference. Thus the very different screening responses do not
correspond to switching an additional localized $4f$ spectator on or
off.
They reflect different rehybridizations of the occupied Ce--O states.

An analogy can be drawn with the effective, single-configuration
ligand-field description of multiplet spectra. In its molecular-orbital
interpretation, the nominal $d^n$ configuration counts electrons in an
active manifold of metal-like nonbonding or antibonding orbitals, while
the filled, predominantly ligand-like bonding orbitals remain implicit.
Hybridization contributes to the splitting of the active levels and
introduces metal $d$ character into the filled bonding states. This
projected metal charge is compatible with an unchanged integer
occupation of the effective multiplet space~\cite{quanty}.

Similarly, the approximately one-electron increase in projected absorber
$4f$ population in the primary CeO$_2$ state, relative to the transported
sudden reference, is carried by occupied hybridized Ce--O states without
the emergence of an additional nearly filled atomic-like $4f$ orbital.
Within the present description, this covalent charge contributes to the
self-consistent screening background rather than constituting an
additional localized multiplet spectator. The effective $4f^0$ label
used for both screening states denotes the absence of such a spectator,
not the absence of atomic $4f$ character in occupied electronic states.
For the initial local sector, the $4f^n$ label counts the
localized electrons retained as spectators in the multiplet
problem, rather than the total atomic-projected $4f$
population of the screening background. Full projected populations,
local-density eigenvalues, and the corresponding-orbital decomposition
are reported in the Supplemental Material.

\subsection{Spectroscopic consequences of the two screening states}
\label{sec:screening-spectra}

\begin{figure*}[t]
 \centering
 \maybegraphics{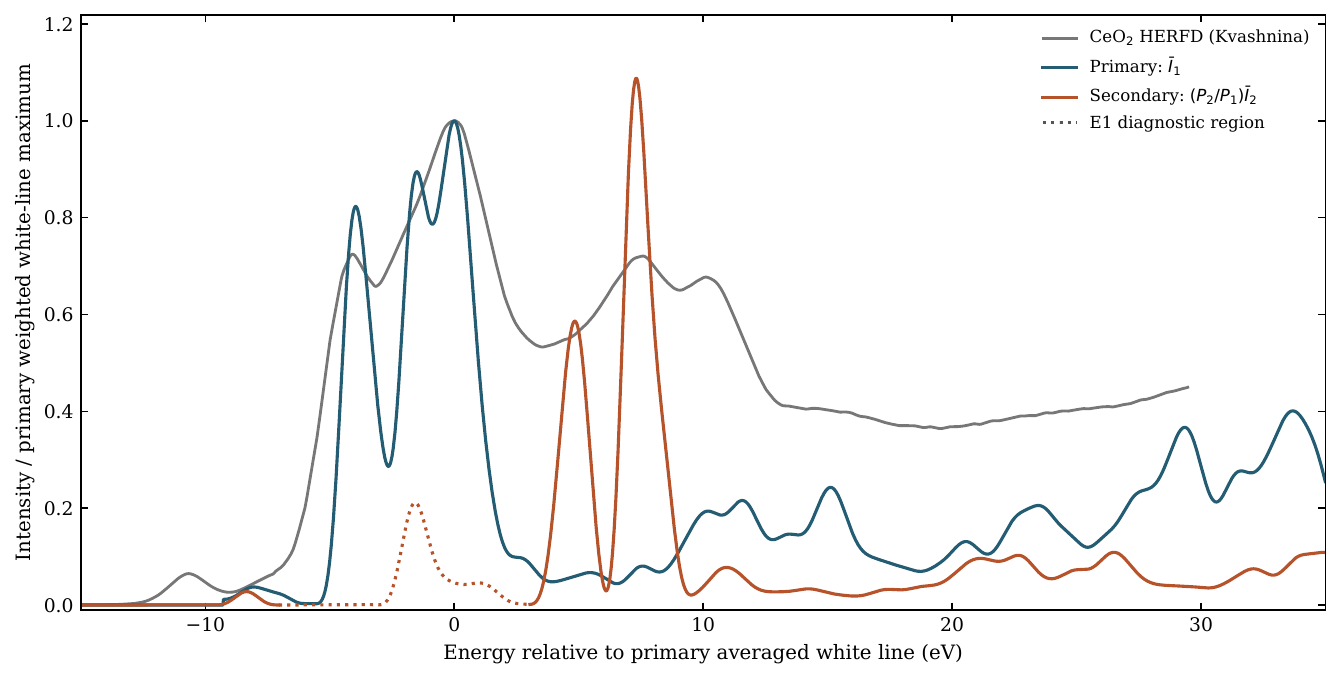}{0.90\textwidth}
 \caption{CeO$_2$ HERFD spectrum and the two calculated screening
 contributions, shown separately. The primary averaged white line sets
 energy zero and weighted unit intensity. With calculated offset $\Delta_2$
 and relative weight $P_2/P_1$, the secondary maximum
 lies 7.31~eV above the primary. Experiment, digitized from
 Ref.~\onlinecite{Kvashnina2024}, is independently centered and normalized
 at its measured white line for qualitative comparison. The dipole response
 is averaged over a $3\times3\times3$ mesh; the remaining contribution is
 approximated by the difference between the combined E1+E2 and E1 responses
 at $\Gamma$. Calculated curves are displayed above a minimum below the
 respective $4f$ pre-edge. The dotted part of the secondary curve covers
 the two E1 features whose dominant diagnostic components are nearly
 fully occupied, together with the intervening region between neighboring
 minima. The dotted continuation is a display convention, not a pointwise
 Pauli classification. Its lower-energy quadrupolar $4f$ resonance remains
 solid, reflecting its local quadrupolar assignment.
 Appendix~\ref{app:spectral-analysis} gives the occupied-state diagnostic
 and averaging prescription. Calculations use 10000 Jackson moments;
 experimental resolution is approximately 0.8~eV HWHM, with no
 fitted additional broadening.}
 \label{fig:spectra}
\end{figure*}

The two screening potentials produce white-line maxima separated by about
7.3~eV, although both calculations retain the same local $4f^0$ sector.
Figure~\ref{fig:spectra} places their contributions on a common energy
scale and compares them with the experimental CeO$_2$ HERFD spectrum
reported by Kvashnina~\cite{Kvashnina2024}. The primary contribution follows
the main white-line structure, whereas the secondary contributes in the
higher-energy satellite region.

This relative placement is determined by the calculation, rather than by
aligning each contribution with an experimental feature. The secondary
charged cell is 11.73~eV higher in QE total energy. On the internal
energy scales of their respective spectral Hamiltonians, however, its
white-line maximum at $\Gamma$ is 4.59~eV lower. Including both terms in
Eq.~\eqref{eq:white-line-accounting} therefore places the secondary white
line 7.13~eV above the primary. Brillouin-zone averaging changes this
separation only slightly, to 7.31~eV.

The relative intensities are likewise not adjusted to the experiment.
Each spectrum retains its calculated transition strength and is multiplied
by the corresponding sudden weight in
Eq.~\eqref{eq:screening-weighted-spectrum}. Both contributions are then
shown on a common scale, with the weighted primary white-line maximum set
to unity. Although the secondary screening weight is smaller,
$P_2/P_1\simeq0.47$, its larger intrinsic peak height gives a weighted
maximum of 1.09 on this scale. The weight of an electronic screening
channel should therefore not be identified with the height of its
white-line peak.

Brillouin-zone averaging affects the higher-energy dipole structure more
strongly than the white-line positions. As shown in
Fig.~\ref{fig:kdependence}, the primary and secondary white-line maxima
shift by only $-0.12$ and $+0.05$~eV, respectively, relative to their
$\Gamma$-point values. In contrast, the higher dipole feature of the
primary channel, near 11.5~eV above its own $\Gamma$ white line,
decreases from about
0.45 to 0.22 in relative peak height upon averaging. This feature belongs
to the primary screening channel and is distinct from the white line
associated with the secondary screening state. The selected resonance
energies and intensities are collected in
Table~\ref{tab:selected-resonances} in
Appendix~\ref{app:spectral-analysis}.

The comparison suggests that distinct collective screening responses can
account for contributions in both the main white line and the satellite
region without introducing an additional localized $4f$ spectator.
The calculated curves include numerical Jackson broadening, but no
additional convolution describing experimental resolution and lifetime
effects. Convolution with a normalized profile would redistribute intensity
while preserving the total spectral area. Peak-height differences alone
therefore do not establish agreement or disagreement in resonance strengths;
a quantitative comparison of resonance areas is not attempted here.

The additional high-energy satellite shoulder is not reproduced by the
two-channel model. It could receive contributions from
an additional stationary screening determinant with appreciable sudden
overlap, or from a family of screening states carrying appreciable
combined weight.

\subsection{Dipole and quadrupole contributions to the pre-edge}
\label{sec:photoelectron-character}

\begin{figure*}[tp]
 \centering
 \maybegraphics{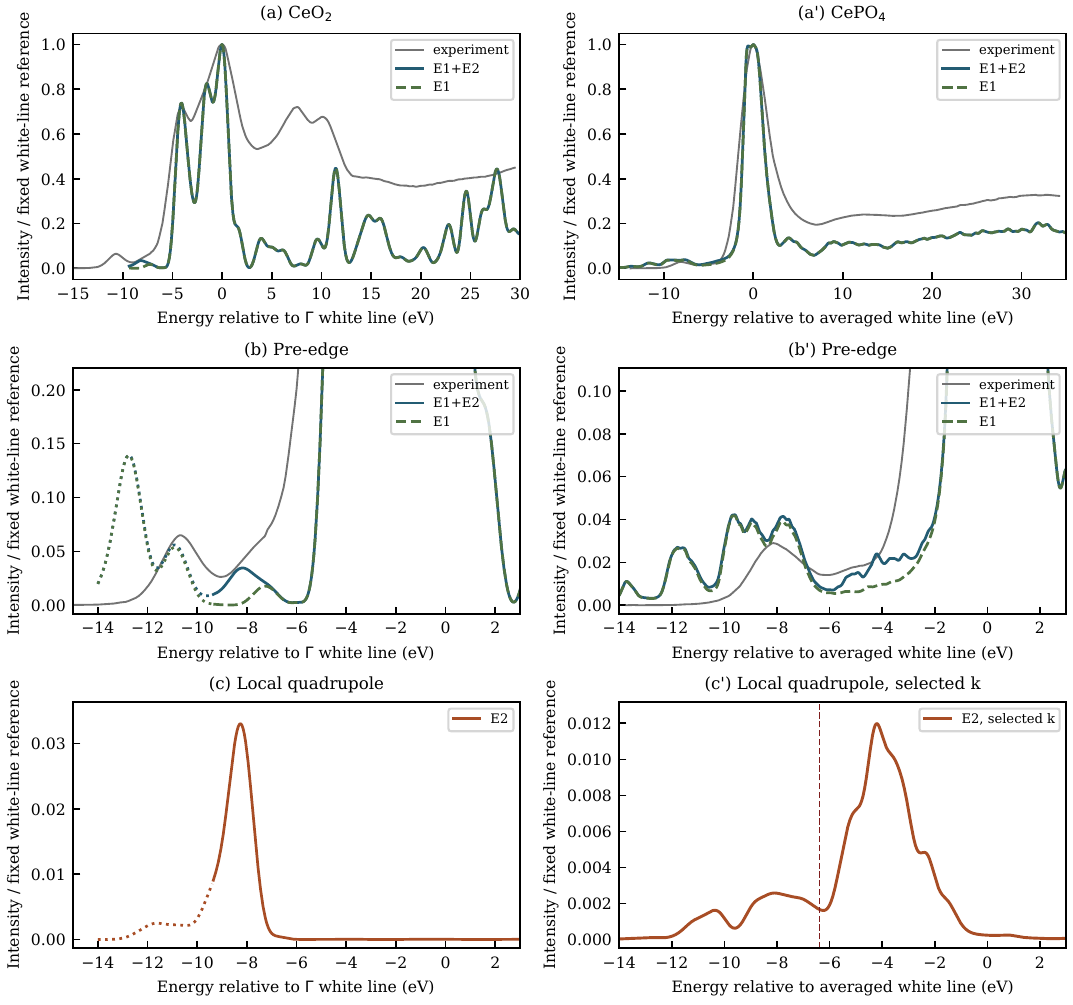}{0.98\textwidth}
 \caption{Dipole and quadrupole contributions to the Ce $L_3$ spectra of
 CeO$_2$ (left) and CePO$_4$ (right). The top row shows the near-edge
 spectra, the middle row enlarges the pre-edge (E1+E2 and E1 in the
 CePO$_4$ panel), and the bottom row resolves the quadrupole response.
 The CeO$_2$ calculation uses the primary
 screening potential at $\Gamma$ with 10000 Chebyshev moments and a
 Jackson kernel. CePO$_4$ E1 and E1+E2 use 3000 Lanczos iterations and
 averages over all 27 points of a shifted $3\times3\times3$ mesh,
 as described in Sec.~S6 of the Supplemental Material. Pure E2 uses 10000
 Jackson--Chebyshev moments at $\bm k=(1/6,1/6,1/6)$.
 Lorentzian HWHM is 0.3~eV in (a') and 0.1~eV in (b',c'), additional
 to Jackson smoothing in (c'). Calculated
 channels retain their relative transition strengths. Energy zero is
 the primary $\Gamma$ white line for CeO$_2$, and the averaged
 E1+E2 white line at HWHM 0.3~eV for all CePO$_4$ panels.
 No relaxation shift is applied to the curves.
 The red dashed line marks the heuristic CePO$_4$ pre-edge estimate
 at $-6.38$~eV. It is obtained by shifting the $-4.13$~eV maximum
 at HWHM 0.3~eV by the transferred correction of $-2.25$~eV,
 at fixed white-line position; it does not use the maximum of the
 HWHM 0.1~eV curve displayed in (c'). The CePO$_4$
 response is shown over the full available energy window, without a
 line-style change in its low-energy tail. The two right-hand
 pre-edge panels share the same energy limits, $[-14,3]$~eV. Dotted
 continuations in the lower-left panels retain the
 CeO$_2$ regions below the cutoff used in Figs.~\ref{fig:kdependence}
 and~\ref{fig:spectra}; they do not classify every point as occupied or empty.
 Gray curves are the HERFD spectra digitized from Fig.~1 of
 Ref.~\onlinecite{Kvashnina2024}, independently aligned and normalized at
 each material's white line. They are shown throughout the plotted
 windows and used only for qualitative comparison. The differing
 numerical broadenings are not used to compare linewidths between the
 two materials.}
 \label{fig:paired-materials}
\end{figure*}

The E1 and E2 decomposition in the left panels of
Fig.~\ref{fig:paired-materials} identifies the dominant CeO$_2$ pre-edge
as an absorber-centered $4f$ resonance excited through the quadrupole
channel. In the primary screening potential, taking its $\Gamma$
white-line maximum as zero, the E2 maximum lies
\begin{equation}
 E_{4f}^{\mathrm{E2}}-E_{\mathrm{WL}}\simeq-8.26~\mathrm{eV}.
 \label{eq:gamma-peaks}
\end{equation}
At this energy the quadrupole intensity is about 0.033 on the white-line
scale, compared with only 0.0013 for the dipole contribution. A separate,
weaker E1 maximum occurs at $-7.25$~eV on the same primary-$\Gamma$
white-line-relative scale, with an intensity of about
0.018. This resonance has $4f$ character on neighboring Ce sites,
rather than on the absorber. It is reached through the dipole-allowed
$l=2$ component at the absorbing Ce, whereas the E2 resonance probes
the absorber's own $4f$ states. The E1/E2 labels therefore specify
the local excitation channel at the absorber, not the orbital character
of the photoelectron on every site. Site- and angular-momentum-resolved
projections supporting this assignment are described in Sec.~S7 of the
Supplemental Material.

The occupied character of a photoelectron component cannot be inferred
from its energy relative to the $4f$ pre-edge alone.
The occupied-state diagnostics use separately constructed wavepackets in
the same screening backgrounds. They provide conditional assignments for
those components, not a projection of the channel-summed intensities shown
here. The energy-filtering procedure and the correspondence with spectral
features are described in
Sec.~S7 of the Supplemental Material.
PAW projections of the selected dominant local components give negligible
occupied weight for the assigned primary pre-edge and white-line resonances,
whereas the lower structures are almost entirely occupied.
For the secondary screening potential at $\Gamma$, taking its own
white-line maximum as zero, the two E1 maxima at approximately
$-8.70$ and $-6.41$~eV have occupied weights 0.999893 and 0.999947.
This energy zero differs from the primary averaged white line used for
the common comparison in Fig.~\ref{fig:spectra}.
Within the secondary spectrum, these structures lie above the local
quadrupolar $4f$ pre-edge. The assignment of this secondary pre-edge is
based on its local quadrupolar character and the nearly empty absorber
$4f$ subspace discussed above, rather than on a direct occupied-space
projection of that packet.
Figures~\ref{fig:kdependence} and \ref{fig:spectra} therefore show the
E1 region flagged by these diagnostics as dotted segments while retaining the $4f$ resonance
as a solid curve wherever the quadrupole contribution is present.
The dotted continuation between minima is a display convention; the
projection measures the dominant local channel at the two selected
energies, rather than every point in that interval.
The diagnostic and the limits of the displayed energy window are described in
Appendix~\ref{app:spectral-analysis} and the Supplemental Material.

\subsection{State-specific relaxation and relative excitation energies}
\label{sec:relaxed-energies}

Within each screening channel, the spectra discussed so far use the same
occupied electronic background for all photoelectron states. They therefore
do not allow that background to respond differently to a localized $4f$
excitation and to the more diffuse white-line state. To assess this
additional relaxation, we construct two neutral final states by injecting
an additional electron into the primary charged screening determinant.
The injected orbitals are obtained from resonance wavepackets centered on
the $4f$ pre-edge and white-line energies, using a Chebyshev energy filter
with Jackson damping~\cite{kernel}. Their photoelectron components are
projected onto the complement of the occupied PAW subspace before injection.

\begin{samepage}
Each neutral determinant is optimized separately, using its initial orbitals
as the fixed state-selection reference.
The construction and optimization of these states are described in
Appendix~\ref{app:state-selection}; numerical parameters, convergence
checks, and total energies are reported in Sec.~S4 of the Supplemental
Material.

\end{samepage}

The difference between their relaxed total energies is
\begin{equation}
 \Delta E_{5d-4f}^{\mathrm{relax}}\simeq10.51~\mathrm{eV},
 \label{eq:relaxed-gap}
\end{equation}
compared with the 8.26-eV separation between the E2 pre-edge and white-line
maxima of the primary fixed-background spectrum at $\Gamma$.
The 2.25-eV difference provides an estimated correction to the
fixed-background gap, including relaxation of both neutral states.
This comparison combines total-energy and spectral descriptions;
it does not define a universal shift of all spectral features.

\subsection{\texorpdfstring{CePO$_4$}{CePO4}: the open-shell counterpart}
\label{sec:cepo4}

Two distinct self-consistent magnetic screening solutions are obtained
for CePO$_4$. The reference used for the spectra retains one nearly filled
local $4f$ direction on the absorber and has a geometric mean squared
overlap of 0.723 with the sudden reference. The alternative contains
a second nearly filled local $4f$ direction, but its corresponding
overlap is only 0.00138.

The search uses staged SCF and fixed-reference KS--IMOM preparations,
followed by ordinary SCF with freely updated Hubbard matrices and
total magnetization constrained to zero. Both solutions use a 24-atom
monazite cell. We call the stronger-overlap solution primary even though
the alternative has a lower smearing free-energy estimate, and calculate
no separate spectrum for the alternative. The geometric overlap
diagnostic uses integer-occupied representations; it is not an intensity
weight for the Gaussian density used in spectroscopy. Preparation,
parameters and numerical checks are given in Sec.~S2.3 of the Supplemental
Material. Figure~\ref{fig:paired-materials} shows Brillouin-zone
averages of E1 and E1+E2, and pure E2 at
$\bm k=(1/6,1/6,1/6)$, all relative to the primary averaged E1+E2
white line at HWHM 0.3~eV.

\FloatBarrier
The primary's local spectroscopic $4f^1$ configuration explicitly
includes a localized spectator electron. Quadrupole excitation therefore
reaches a local $2p^54f^2$ manifold, rather than the $2p^54f^1$ manifold
of CeO$_2$. The spectator participates directly in the multiplet
interactions with the core hole and the photoelectron.

At the selected wave vector, the 10000-moment Jackson--Chebyshev response
with additional Lorentzian HWHM 0.3~eV has its principal quadrupole
pre-edge 4.13~eV below this white line.
The combined response also contains dipole intensity in the pre-edge region;
the local quadrupole multiplet does not exhaust the pre-edge intensity.
The narrower spectra in Fig.~\ref{fig:paired-materials}(b',c') resolve
the structured pre-edge. Lanczos and Chebyshev agree when compared with
the same spectral kernel; the finite resolution of the expansion remains
relevant to the finest features. The recurrence and momentum checks,
and the different convergence behavior of the two
recursions~\cite{Mirone2026} are discussed in the Supplemental Material.

As in CeO$_2$, the low-energy tail cannot be assigned from its energy or
shape alone. For the dominant local components examined in CePO$_4$,
the occupation-weighted overlaps indicate mixed character in the lower
tail and predominantly unoccupied character closer to the pre-edge.
The sampled components do not define a sharp
occupied-state boundary or a channel-summed blocked intensity. We
therefore retain the full calculated CePO$_4$ response without using
line-style changes to classify its low-energy tail as allowed or blocked.
The conditional projection is defined in
Appendix~\ref{app:spectral-analysis}; the numerical weights are given
in the Supplemental Material.

For CePO$_4$, we transfer the estimated CeO$_2$ gap correction:
the 10.506-eV difference between relaxed neutral-state total energies
exceeds the 8.256-eV fixed-background spectral gap by 2.250~eV.
Assuming that this net correction is comparable in CePO$_4$, we lower the $4f$
multiplet response by this amount relative to the white line, without
changing its internal structure. The estimated gap is therefore
\begin{equation}
 G_{\mathrm{CePO_4}}^{\mathrm{est}}
 \simeq (4.13+2.25)~\mathrm{eV}=6.38~\mathrm{eV}.
 \label{eq:cepo4-estimated-peak}
\end{equation}
This places the principal peak at $-6.38$~eV relative to the calculated
white line, about 1.7~eV above experiment ($-8.1$~eV from its measured
white line). The transfer is heuristic; no pair of relaxed neutral-state
energies has been calculated for the current CePO$_4$ reference.
The transfer is defined in Appendix~\ref{app:state-selection}.

\section{Conclusions and outlook}
\label{sec:conclusions}

The Ce $L_3$ near-edge structure reflects the interplay between local
multiplet dynamics, collective core-hole screening, and photoelectron
propagation. For Ce$^{4+}$ in CeO$_2$, our heuristic search identifies
primary and secondary stationary screening determinants whose squared
overlaps with the common sudden reference give approximate sudden weights
of 63\% and 30\%, respectively. Both retain the same local spectroscopic
$4f^0$ sector, while their white-line maxima are separated by about
7.3~eV. The two calculated contributions account for features in both
the main white-line and satellite regions. The remaining
features could involve a third determinant with appreciable sudden
overlap, or a family of screening states carrying appreciable combined
weight.

For Ce$^{3+}$ in CePO$_4$, the primary self-consistent magnetic screening
state has a geometric mean squared overlap of 0.723 with the sudden
reference across the sampled $\bm k$ points, using the integer-occupation
convention. A second stationary solution has stronger local $4f$
occupation on the absorber and very little overlap with the transported
reference. It is not included as a second spectral contribution.
The contrast reinforces the use of overlap diagnostics alongside
stationarity and total energy when identifying screening channels.
The primary's local $4f^1$ spectator explicitly enters
the multiplet dynamics and gives rise to the $4f^2$ quadrupole
final-state manifold.

In the fixed-background XSpectruplet calculations, the principal $4f$
E2 resonance in CeO$_2$ lies 8.26~eV below the primary $\Gamma$ white line.
The selected CePO$_4$ resonance lies 4.13~eV below its averaged
E1+E2 white line. Independent neutral-state relaxation gives a CeO$_2$
gap of 10.51~eV, close to the experimental value of about 10.7~eV.
Its excess of 2.25~eV over the fixed-background spectral gap defines
the net correction. Transferring this correction heuristically to
CePO$_4$, without changing its internal multiplet structure, gives an
estimated gap of 6.38~eV, still about 1.7~eV below the experimental
separation of approximately 8.1~eV~\cite{Kvashnina2024}.
Table~\ref{tab:selected-resonances} distinguishes the explicit CeO$_2$
relaxation result from the transferred CePO$_4$ estimate and compares
both with the fixed-background and experimental positions.

These aspects of the excitation must be treated within a consistent
theoretical framework. XSpectruplet couples the local multiplet and the
continuum photoelectron in a single spectral description. Using
optimized Kohn--Sham electronic backgrounds provides access to distinct
screening responses and their spectroscopic signatures. For CeO$_2$,
within the independent-channel approximation, their overlaps with the
sudden reference provide approximate weights for combining the
corresponding spectra.

The present heuristic search begins to unravel how these ingredients
shape the Ce $L_3$ edge. Further progress will require more systematic
algorithms for locating additional stationary solutions of the
Kohn--Sham equations with non-negligible overlap with the electronic
state prepared in the sudden approximation. Such developments would extend
the search for screening states and allow us to assess how far additional
independent screening contributions account for the remaining near-edge
structure.

\appendix
\section{PAW transport and excitation-source construction}
\label{app:paw-source}

Let $T_Z$ and $T_{Z+1}$ be the PAW transformations associated with the
ground-state and equivalent-core atomic inputs of the absorber. Their
overlap operators are
\begin{equation}
 S_Z=T_Z^\dagger T_Z,\qquad
 S_{Z+1}=T_{Z+1}^\dagger T_{Z+1}.
\end{equation}
The occupied ground-state orbitals are transported into the
equivalent-core representation by solving
\begin{equation}
 S_{Z+1}C_{Z\rightarrow Z+1}
 =T_{Z+1}^\dagger T_Z C_Z.
 \label{eq:paw-transport}
\end{equation}
We use an orthonormal basis of the transported occupied subspace in the
target PAW metric. This change of basis leaves the occupied subspace and
its normalized Slater determinant unchanged; it involves no orbital
relaxation or minimization of the target Hamiltonian. Numerical details
of the transport and orthonormalization are given in the Supplemental
Material.

The excitation source is constructed separately. For multipole rank $K$
and final angular channel $L$, its radial form is
$u_L^{\mathrm{target}}(r)=r^K u_c(r)$, where $u_c(r)$ is the radial core
orbital. For each angular channel, we consider all-electron radial
eigenfunctions inside a spherical box of radius $R_{\mathrm{box}}$.
The finite box provides
a discrete set of radial states in which to expand the excitation source.

The cutoff refers to eigenvalues of this auxiliary atomic radial problem,
not photon transition energies. Retaining states with
$E_j\le E_{\mathrm{cut}}^{\mathrm{radial}}$, we write
\begin{equation}
 u_L^{\mathrm{target}}(r)
 \simeq \sum_j c_{jL}u_{jL}^{\mathrm{AE}}(r).
 \label{eq:radial-map}
\end{equation}
For each excitation channel, let $|b_{\rm AE}\rangle$ denote the
all-electron source obtained from the radial expansion in
Eq.~\eqref{eq:radial-map}, after applying the radial window and including
the angular and spin factors. We seek its representation
$|\widetilde b\rangle$ in the retained plane-wave PAW space. The fixed
transformation $\mathcal T$ supplied by the equivalent-core pseudopotential
maps a pseudo function $|\widetilde\psi\rangle$ to its all-electron form
$\mathcal T|\widetilde\psi\rangle$. With $\mathcal P_{\rm PW}$ denoting
the projector onto the retained plane waves, the metric in this space is
$S_{\rm PW}=\mathcal P_{\rm PW}\mathcal T^\dagger\mathcal T\mathcal P_{\rm PW}$.

The source representation must preserve its overlap with every test
function $|\widetilde\psi\rangle$ in that space:
\[
 \begin{aligned}
 \langle\widetilde\psi|S_{\rm PW}|\widetilde b\rangle
 &=\langle\mathcal T\widetilde\psi|b_{\rm AE}\rangle\\
 &=\langle\widetilde\psi|\mathcal T^\dagger|b_{\rm AE}\rangle.
 \end{aligned}
\]
Consequently, $S_{\rm PW}|\widetilde b\rangle
=\mathcal P_{\rm PW}\mathcal T^\dagger|b_{\rm AE}\rangle$, giving
\begin{equation}
 |d\rangle=\mathcal P_{\rm PW}\mathcal T^\dagger|b_{\rm AE}\rangle,
 \qquad
 |\widetilde b\rangle=S_{\rm PW}^{-1}|d\rangle,
\end{equation}
Here $|d\rangle$ contains the ordinary plane-wave projections of
$\mathcal T^\dagger|b_{\rm AE}\rangle$. The inverse metric converts them
into the PAW source vector; it is not a scalar normalization.
The inverse metric is applied in that plane-wave space before the local
channel exclusions and equivalent-placement completion.

Increasing the radial-energy cutoff includes more rapidly oscillating
states in the all-electron expansion; it does not change the PAW
transformation or introduce independently pseudized box states. We use
$E_{\mathrm{cut}}^{\mathrm{radial}}=20$~Ry for the spectra reported here.
The radial-basis parameters and cutoff tests are given in the
Supplemental Material.

\section{State selection and relaxed-state calculations}
\label{app:state-selection}

\subsection{Neutral final states of \texorpdfstring{CeO$_2$}{CeO2}}

In KS--IMOM, the spin-resolved KS Hamiltonian is diagonalized at
each SCF iteration, and occupied states are selected by their
PAW-metric overlap with a fixed reference~\cite{Gilbert2008,Barca2018}.
The orbital overlaps used for this selection are distinct from the
many-electron determinant overlaps evaluated after optimization.
State tracking, eigenproblem accuracy, and charge-density convergence
are monitored separately.

The two neutral CeO$_2$ final states are constructed from the same
primary charged screening determinant. To obtain an initial orbital for
each added electron, we first form a resonance wavepacket around the
selected $4f$ pre-edge or white-line energy. An energy filter is applied
to the excitation source through a finite Chebyshev expansion of the
spectral Hamiltonian. Jackson damping suppresses truncation oscillations
and gives the energy filter an approximately Gaussian profile~\cite{kernel}.
The photoelectron component in the dominant local channel of the packet
provides the orbital for injection.

This orbital is projected onto the complement of the occupied PAW
subspace, normalized, and added in the appropriate spin channel. The
density, PAW augmentation, and Hubbard occupation matrix are reconstructed
for the resulting 769-electron determinant. Each neutral state is then
optimized independently. Its initial occupied orbitals remain the fixed
reference for IMOM state selection, while the electronic density and
Hubbard occupation matrix are updated during the SCF optimization.

The reported white-line optimization uses 1000 KS bands to represent
the diffuse wavepacket; a 472-band space does not represent it adequately.
The optimized determinant retains a squared overlap of about 0.4 with
its initial neutral reference. The relaxed white-line--$4f$ total-energy
separation is 10.506~eV, with a variation of about 1~meV in the $4f$-state
energy upon a further Hubbard-matrix update. QE total cell energies,
convergence criteria, and the Hubbard-update procedure are reported
in the Supplemental Material.

\subsection{Transferred gap estimate for
\texorpdfstring{CePO$_4$}{CePO4}}

Let $G_M^{\mathrm{spec}}$ be the positive white-line--$4f$ peak
separation in the fixed-background spectrum of material $M$.
For CeO$_2$, the independently relaxed neutral-state total energies
give $G_{\mathrm{CeO_2}}^{\mathrm{relax}}=10.506$~eV, compared with
$G_{\mathrm{CeO_2}}^{\mathrm{spec}}=8.256$~eV at $\Gamma$.
We define the estimated gap correction by comparing these two quantities:
\begin{equation}
 \Delta_{\mathrm{est}}
 = G_{\mathrm{CeO_2}}^{\mathrm{relax}}-G_{\mathrm{CeO_2}}^{\mathrm{spec}}
 \simeq 2.250~\mathrm{eV}.
\end{equation}
The total-energy gap uses both optimized neutral endpoints. The comparison
with the fixed-background spectral gap estimates the state-specific
correction to that gap using two different energy descriptions.
We transfer this net correction heuristically to CePO$_4$:
\begin{align}
 G_{\mathrm{CePO_4}}^{\mathrm{est}}
 &\simeq G_{\mathrm{CePO_4}}^{\mathrm{spec}}+\Delta_{\mathrm{est}}
 \nonumber\\
 &\simeq (4.13+2.25)~\mathrm{eV}=6.38~\mathrm{eV}.
\end{align}
The CePO$_4$ spectral gap uses the E2 maximum at
$\bm k=(1/6,1/6,1/6)$ relative to the primary averaged
E1+E2 white line. Both use Lorentzian HWHM 0.3~eV; the E2 curve also
retains its M10000 Jackson kernel.
Thus the estimated peak lies at $-6.38$~eV on this white-line-relative
axis. No absolute photon energy is transferred between materials.

This estimate assumes comparable net gap corrections
and treats the $4f$ multiplet as translated without changing its internal
structure. It is not obtained from a new pair of relaxed neutral
CePO$_4$ states.

\section{Computational parameters and numerical controls}
\label{app:parameters}

CeO$_2$ is represented by a 96-atom fluorite cell with one absorber
described by the full-$Z+1$ PAW atomic input, cell charge $+1$, and
768 electrons ($384\uparrow+384\downarrow$). The PBE+$U$ functional uses
the Dudarev form with $U_{\mathrm{eff}}=5$~eV on every Ce $4f$ manifold
\cite{pbe,Dudarev1998}. Wavefunction and density cutoffs are 103 and
1100~Ry. Calculations at $\Gamma$ use complex wavefunctions.
The CeO$_2$ spectral expansion contains 10000 Chebyshev moments over
$[-63.5,165]$~Ry on the internal XSpectruplet Hamiltonian scale, before
white-line alignment, and uses a Jackson kernel. The Slater integrals are
85\% of their atomic values, and the Ce $2p$ spin--orbit interaction is
scaled by 0.2. This is a numerical compromise for the $L_3$-focused
calculation: stronger core spin--orbit splitting widens the spectral
range to be represented by polynomial recursions. In particular, a wider
energy interval requires more Chebyshev moments to maintain a given
energy resolution. Sensitivity to this choice is assessed in Sec.~S5
of the Supplemental Material.
The E1 Brillouin-zone average uses eight representatives of
an unshifted $3\times3\times3$ mesh, with multiplicities summing to 27.

CePO$_4$ uses the 24-atom monazite cell~\cite{Ni1995}, a charged
164-electron reference, and the same PBE+$U$ functional and plane-wave
cutoffs. E1 and E1+E2 spectra are averaged over all 27 points of a shifted
$3\times3\times3$ mesh, as specified in Sec.~S6 of the Supplemental
Material. E2 is evaluated at $(1/6,1/6,1/6)$.
The local initial sector is $4f^1$, the Slater
integrals are 75\% of their atomic values, the $4f$ spin--orbit interaction
is retained at full strength, and the Ce $2p$ spin--orbit scale is 0.2.
The magnetic screening reference is obtained by staged searches followed
by ordinary SCF with freely updated Hubbard occupation matrices, reaching
the requested residual threshold of $1.4\times10^{-4}$~Ry.
The final matrices are reconstructed independently from the collected
wavefunctions and checked against the saved spectral potential.
The fractional density used for spectroscopy and the integer filling
used for determinant overlaps are distinguished in the Supplemental
Material, which reports preparation, convergence and local-matrix checks.

CePO$_4$ E1 and E1+E2 use 3000 Lanczos iterations and Lorentzian HWHM
0.3~eV for the overview and 0.1~eV for the detail. Their means
differ from the corresponding 1000-iteration results by peak-normalized
RMS values of 0.12--0.16\% and 0.83--1.17\%, respectively.
The displayed E2 response uses 10000 Jackson--Chebyshev moments over
$[-22,118]$~Ry and the stated additional Lorentzian broadening.
The selected-point Lanczos and Chebyshev responses agree within
$6\times10^{-7}$ in relative $L^2$ norm over the pre-edge after applying
the same kernel. Their finite-resolution and sampling checks are
reported in the Supplemental Material.

The Supplemental Material documents the E1+E2 sensitivity to the radial-source
cutoff and core spin--orbit scale, recurrence-length comparisons, moment
bounds, and reference-state diagnostics used to assess the numerical
stability of the reported spectra.

\section{Spectral averages and occupied-state diagnostics}
\label{app:spectral-analysis}

The Brillouin-zone average combines the spectra while preserving their
calculated transition strengths. For E1 polarization along $x$, the
CeO$_2$ average is
\begin{equation}
 I_{\mathrm{avg}}(E)=\sum_r\frac{m_r}{27}
 \lVert b_r\rVert_S^2 I_r^{\mathrm{normalized}}(E),
 \label{eq:k-average}
\end{equation}
Here $m_r$ is the multiplicity of representative $r$.
The response $I_r^{\mathrm{normalized}}$ is calculated for a unit-norm
excitation source, so multiplication by $\lVert b_r\rVert_S^2$ restores
its transition strength. The individual spectra are combined on their
common energy scale without independent peak alignment or peak-height
normalization.

For the two-channel CeO$_2$ comparison in Fig.~\ref{fig:spectra}, the
combined response of screening state $j$ is approximated by
\begin{align}
 \bar I_j(\epsilon)\simeq{}&
 \bigl\langle I_j^{\mathrm{E1}}(\epsilon,\bm{k})\bigr\rangle_{\bm{k}}
 \nonumber\\
 &+I_j^{\mathrm{E1+E2}}(\epsilon,\Gamma)
 -I_j^{\mathrm{E1}}(\epsilon,\Gamma).
 \label{eq:approximate-multipole-average}
\end{align}
The difference between the combined and dipole responses at $\Gamma$
is held fixed across the mesh. This neglects its momentum dependence;
it is not an independently calculated E2 Brillouin-zone average.
All terms use the same Jackson expansion. Reference-state energy shifts
are then included as in Eq.~\eqref{eq:screening-weighted-spectrum}.
Using the unrounded determinant overlaps gives
$P_2/P_1\simeq0.47$ and a weighted white-line peak-height
ratio of approximately 1.09.
The experimental trace
is independently aligned and normalized and is not used to fit these
weights.

\begin{table*}[t]
 \caption{(a) Selected resonances of the primary CeO$_2$ screening potential
 at $\Gamma$, with energies and peak intensities relative to its
 $\Gamma$ white-line maximum.
 The local angular channel identifies the excitation at the absorber.
 (b) E2 pre-edge positions relative to the corresponding white-line
 reference for each material and treatment; all energies are in eV.
 Fixed-background positions use $\Gamma$ for CeO$_2$ and the
 selected-$\bm{k}$ Jackson--Chebyshev E2 calculation relative to the
 averaged E1+E2 white line, with additional Lorentzian
 HWHM 0.3~eV, for CePO$_4$.
 The CeO$_2$ relaxation estimate is the negative of the independently
 relaxed neutral-state energy separation. The CePO$_4$ estimate adds
 the heuristically transferred CeO$_2$ gap correction of 2.25~eV, as described in
 Appendix~\ref{app:state-selection}. Experimental positions are relative
 to each material's measured white line and are taken from
 the HERFD comparison with Ref.~\onlinecite{Kvashnina2024}. The CeO$_2$
 and CePO$_4$ values are graphical estimates from the digitized HERFD
 curves, reported to one decimal place.
 Relaxation estimates specify energies only, not recalculated intensities.}
 \label{tab:selected-resonances}
 \begin{ruledtabular}
 \begin{tabular}{llccc}
 \multicolumn{5}{c}{(a) Fixed-background CeO$_2$ spectral resonances} \\
 Resonance & Excitation & $E-E_{\mathrm{WL}}$ & $I/I_{\mathrm{WL}}$
 & Local angular channel \\
 \hline
 Absorber $4f$ pre-edge & E2 & $-8.26$ & 0.033 & $l=3$ \\
 Neighboring-Ce $4f$ pre-edge & E1 & $-7.25$ & 0.018 & $l=2$ \\
 White line & E1+E2 & 0 & 1 & predominantly $l=2$ \\
 Higher dipole resonance & E1 & $+11.45$ & 0.45 & $l=2$ \\
 \hline
 \multicolumn{5}{c}{(b) E2 pre-edge position: calculation, relaxation, and experiment} \\
 \multicolumn{2}{l}{Material} & Fixed background & \shortstack{Relaxation-based\\estimate} & HERFD experiment \\
 \hline
 \multicolumn{2}{l}{CeO$_2$} & $-8.26$ & $-10.51$ & $\sim -10.7$ \\
 \multicolumn{2}{l}{CePO$_4$} & $-4.13$ & $-6.38$ (transferred correction) & $\sim -8.1$ \\
 \end{tabular}
 \end{ruledtabular}
\end{table*}

To distinguish absorption resonances from contributions associated with
occupied background states, we analyze the energy-filtered wavepackets
obtained by the energy-filtering step described in
Appendix~\ref{app:state-selection}. In the spectral representation of
Eq.~\eqref{eq:spectral-representation}, the filter retains the eigenstates
contributing near the selected energy, weighted by their excitation
amplitudes. For an isolated resonance dominated by one eigenstate, the
normalized packet has essentially that state's character; for overlapping
levels it characterizes their source-weighted combination at the filter's
resolution. The relation to the spectral density and the association of
the diagnostic packets with spectral features are detailed in Sec.~S7
of the Supplemental Material. The occupied-state diagnostic is
evaluated on the photoelectron component before removal of its overlap
with occupied KS orbitals. For packets selected to construct neutral
states, the occupied component is removed and the remainder normalized
before injection; this is a separate operation.
For the photoelectron component of a selected packet in the dominant local channel
$\alpha_*$, at fixed spin $\sigma$ and wave vector $\bm k$, the occupied
weight is evaluated as
\begin{equation}
 W_{\mathrm{occ}}^{(\alpha_*)}=
 \frac{\sum_n f_{n\bm{k}\sigma}
 |\langle u_{n\bm{k}\sigma}|S|\phi_{\lambda\alpha_*}\rangle|^2}
 {\langle\phi_{\lambda\alpha_*}|S|\phi_{\lambda\alpha_*}\rangle},
 \label{eq:occupied-weight}
\end{equation}
where $u_{n\bm{k}\sigma}$ and $f_{n\bm{k}\sigma}$ are the background
KS orbitals and their occupations in that spin and wave-vector block.
For integer occupations this is a projection onto the occupied
subspace; with fractional occupations it is the expectation of the
occupation operator, which is not idempotent.

\begin{samepage}
For the diagnostic packets specified in Sec.~S7 of the Supplemental
Material, the four primary resonances in Table~\ref{tab:selected-resonances}(a)
have occupied weights between
$2.4\times10^{-7}$ and $4.4\times10^{-5}$. Two lower structures in each
CeO$_2$ screening potential instead have weights between 0.999962 and
0.999999. The diagnostic therefore distinguishes the selected physical
resonances from lower features whose examined photoelectron components
belong almost entirely to the occupied subspace.

\end{samepage}

The combined CeO$_2$ curves retain a lower display cutoff at a minimum
below the $4f$ resonance of each background, located using its quadrupole
response at $\Gamma$. The secondary E1 filters at $\Gamma$, at internal
XSpectruplet energies
$-25.52082$ and $-23.22368$~eV additionally give occupied weights
0.999893 and 0.999947. The corresponding maxima in the E1 average are
$-8.88$ and $-6.32$~eV relative to the secondary averaged white line.
They are displayed as
dotted segments, including the intervening region.
For E1 alone the dotted continuation extends to the left axis limit.
For the combined response it stops on the low-energy side at the minimum
following the quadrupolar $4f$ peak, so that this assigned local resonance
remains solid. The two delimiting minima are $-14.35$ and $-4.28$~eV
relative to the secondary averaged white line, while its $4f$ maximum is
at $-15.65$~eV. All style boundaries are defined on the internal energy
axis, then converted to the common plotted scale by adding the
reference-state offset and subtracting the primary averaged white line.
They do not implement
an exact Pauli projection of the full spectrum. Individual
occupied-state projections are reported in the Supplemental Material.

For CePO$_4$, the same diagnostic was applied at
$\bm{k}=(1/6,1/6,1/6)$ using the fractional occupations of the selected
SCF reference. Each weight concerns the dominant local component at
the selected energy; the values and occupation conventions are reported
in the Supplemental Material. No sum over all local many-body channels
is performed. The CePO$_4$ curves in Fig.~\ref{fig:paired-materials}
are shown throughout the available energy window, without an occupied-state
mask or a line-style threshold in the low-energy tail. Experimental traces
are likewise retained throughout the plotted energy window.

\FloatBarrier
\begin{acknowledgments}
We are grateful to Francesco Mauri for an interesting discussion of
Koopmans-compliant potentials during the early stages of the method's conception.
We thank Alessandro Longo, Christoph Sahle, Mauro Rovezzi,
Lucia Amidani, and Marius Retegan for stimulating discussions
on the interpretation of Ce $L_3$ x-ray absorption spectra
and the challenges of their theoretical description.
The calculations were performed using ESRF computing resources.
\end{acknowledgments}

\bibliographystyle{apsrev4-2}
\bibliography{ce4_l3_refs}

\end{document}


\maketitle
\thispagestyle{fancy}
\FloatBarrier
\section{Scope and computational settings}
\label{sec:settings}
This supplement documents the electronic references, determinant overlaps, local orbital analysis, numerical controls, and energy comparisons that support the main article. Charged screening determinants and neutral photoelectron-injected states are distinguished throughout. The former define the background potentials used for spectroscopy; the latter are optimized to estimate state-specific relaxation.

The electronic calculations use PBE with the rotationally invariant Dudarev correction, $U_{\mathrm{eff}}=5$~eV on every Ce $4f$ subspace, including the absorber~\cite{PBE,Dudarev}. Quantum ESPRESSO supplies the PAW electronic backgrounds, and XSpectruplet~\cite{xspectruplet} couples the local multiplet to the plane-wave photoelectron. The underlying Quantum ESPRESSO version used for these calculations is 7.4.1.

\begin{table}[H]
\centering\small
\caption{Electronic and spectral parameters used for the main comparisons.
The electron counts refer to charged screening cells before photoelectron
injection. Chebyshev bounds are on the internal XSpectruplet Hamiltonian
scale, before white-line alignment; the radial-source cutoff refers
instead to the auxiliary atomic eigenproblem of Sec.~\ref{sec:source}.}\label{tab:settings}
\begin{tabular}{p{0.35\linewidth}p{0.28\linewidth}p{0.28\linewidth}}
\toprule
Quantity & \ceo & \cep\\\midrule
Structure and cell & Fluorite; 96 atoms & Monazite; 24 atoms\\
Composition & 32 Ce, 64 O & 4 Ce, 4 P, 16 O\\
Absorber and cell charge & One full-$Z+1$ Ce; $+1$ & One full-$Z+1$ Ce; $+1$\\
Explicit electron count & 768; $384\uparrow+384\downarrow$ & 164, spin resolved\\
Wavefunction / density cutoff & 103 / 1100~Ry & 103 / 1100~Ry\\
Spectral Brillouin-zone mesh & Unshifted $3\times3\times3$; eight E1 representatives & E1/E1+E2: all 27 shifted $3\times3\times3$ points; E2: $(1/6,1/6,1/6)$\\
Local initial configuration & $4f^0$ & $4f^1$\\
Slater-integral scale & 0.85 & 0.75\\
Core $2p$ spin--orbit scale & 0.2 & 0.2\\
Radial-source energy cutoff & 20~Ry & 20~Ry\\
Main spectral recurrence & 10000 Chebyshev moments & E1/E1+E2: L3000; E2: M10000 Chebyshev\\
Broadening & Jackson kernel & Lorentzian HWHM 0.3 / 0.1~eV; E2 also uses Jackson\\
Chebyshev energy bounds & $[-63.5,165]$~Ry & $[-22,118]$~Ry for selected-$\bm{k}$ tests\\
\bottomrule
\end{tabular}
\end{table}

For \ceo, the absorber's 12-valence-electron Ce PAW input is replaced by a 13-valence-electron equivalent-core input. Keeping the explicitly treated electron count unchanged gives the positively charged screening cell. Calculations at $\Gamma$ use complex wavefunctions. For \cep, the monazite structure has $a=6.7880$~\AA, $b=7.0163$~\AA, $c=6.4650$~\AA, and $\beta=103.43^\circ$~\cite{Ni}. Its local $4f$ spin--orbit interaction is retained at full strength.
The actual CeO$_2$ cell vectors are
$(10.982324325,0,0)$, $(0,10.982324325,0)$ and
$(0,0,10.979780176)$~\AA. Its 32 Ce fractional positions are
$(\bm b+\bm t)/2$, where
\[
 \bm b\in\{(0,0,0),(0,\tfrac12,\tfrac12),
             (\tfrac12,0,\tfrac12),(\tfrac12,\tfrac12,0)\},
 \qquad \bm t\in\{0,1\}^{3}.
\]
The 64 O fractional positions are all triples from
$\{1/8,3/8,5/8,7/8\}^{3}$. The absorber is the Ce at the origin.

The CePO$_4$ Cartesian cell is
\[
 \bm a=(6.788,0,0),\quad \bm b=(0,7.0163,0),\quad
 \bm c=(-1.501542901195,0,6.288210708609)\ \text{\AA}.
\]
The following asymmetric fractional coordinates generate the 24 atoms:
\begin{center}\small
\begin{tabular}{lrrr}\toprule
Site & $x$ & $y$ & $z$\\\midrule
Ce & 0.28182 & 0.15914 & 0.10008\\
P  & 0.30470 & 0.16350 & 0.61240\\
O1 & 0.25080 & 0.00550 & 0.44580\\
O2 & 0.38110 & 0.33200 & 0.49820\\
O3 & 0.47450 & 0.10540 & 0.80420\\
O4 & 0.12680 & 0.21640 & 0.71080\\\bottomrule
\end{tabular}
\end{center}
Apply, modulo unit translations, the four operations
$(x,y,z)$, $(\tfrac12-x,\tfrac12+y,\tfrac12-z)$,
$(-x,-y,-z)$, and $(\tfrac12+x,\tfrac12-y,\tfrac12+z)$.
The first Ce is the absorber; the Ce spin seeds follow $+,-,+,-$ in this
order.
We label these four sites Ce1--Ce4 in the same order, with Ce1 denoting
the absorber. The shortest periodic distances from the absorber to
Ce2, Ce3 and Ce4 are 4.079, 4.149 and 4.301~\AA, respectively.
Ce3, at fractional coordinates $(0.71818,0.84086,0.89992)$, is the
same-spin Ce neighbor in this antiferromagnetic pattern; Ce2 and Ce4
have the opposite majority spin.

The scalar-relativistic PBE PAW inputs have valence electron counts
12 (Ce), 13 (equivalent-core absorber), 6 (O) and 5 (P).
Their configurations are
$5s^25p^64f^15d^16s^{1.5}6p^{0.5}$,
$5s^25p^64f^15d^36s^1$,
$2s^22p^4$, and $3s^23p^3$, respectively.
The two Ce inputs have eight radial projectors with pseudization radii
$(r_s,r_p,r_d,r_f)=(1.3,1.4,1.5,0.9)$~bohr, augmentation radius
1.9~bohr, and core-correction radius 0.8~bohr.
O has four projectors, pseudization radii 1.0/0.9~bohr for $s/p$,
augmentation radii 1.30/1.35~bohr, and core radius 0.7~bohr.
P has six projectors, pseudization radius 1.5~bohr,
augmentation radius 1.8~bohr, and core radius 1.5~bohr.
All use Troullier--Martins pseudization and the PSQ augmentation form.

CeO$_2$ reference determinants use a complex $1\times1\times1$ unshifted
mesh and integer occupations. The neutral magnetic CePO$_4$ reference
uses the shifted $3\times3\times3$ mesh and Marzari--Vanderbilt smearing
of 0.02~Ry; the selected charged reference uses the same mesh with
Gaussian smearing of 0.02~Ry. The half-grid shift gives coordinates
$1/6,1/2,5/6$ along each reciprocal axis. The spectral polarization is
$\boldsymbol\epsilon=\hat{\bm x}$ with photon propagation along
$\hat{\bm z}$ in the Cartesian cell frame; the E1 and E2 components use
the same geometry.
These directions refer to the spectra in the main figures; the geometries
used to initialize the injected wavepackets are specified in
Sec.~\ref{sec:relax}.

\FloatBarrier
\section{Screening-state preparation and determinant overlaps}
\label{sec:overlaps}
\subsection{Sudden reference and PAW normalization}
Let $T_Z$ and $T_{Z+1}$ reconstruct all-electron orbitals from the two PAW representations, with $S_Z=T_Z^\dagger T_Z$ and $S_{Z+1}=T_{Z+1}^\dagger T_{Z+1}$. The columns of $C_Z$ are the occupied ground-state pseudo orbitals; $C_t$ contains their transported counterparts. Transport into the equivalent-core representation is defined by
\begin{equation}
 S_{Z+1}C_t=T_{Z+1}^\dagger T_Z C_Z.
\end{equation}
An orthonormal basis of the transported occupied subspace is obtained from its Gram matrix,
\begin{equation}
 G=C_t^\dagger S_{Z+1}C_t,\qquad C_{\mathrm{sud}}=C_tG^{-1/2}.
\end{equation}
This symmetric orthonormalization leaves the occupied subspace, and hence its normalized Slater determinant, unchanged. It is not an orbital relaxation. An exact transport preserving the reconstructed all-electron orbitals would give $G=I$; the orthonormalization addresses the deviations of the finite numerical representation. The numerical checks give deviations of order $1.41\times10^{-3}$ before orthonormalization and $6.66\times10^{-15}$ afterwards.
Here the residual is the largest absolute matrix element,
$\epsilon_{\mathrm{occ}}=\max_{\sigma,ij}|(G_\sigma-I)_{ij}|$,
rather than a Frobenius norm. For the 700-band transport used for the
primary overlap, the occupied-subspace residual before orthonormalization
is $1.409016\times10^{-3}$;
the maximum occupied--virtual entry is $2.415498\times10^{-4}$.
The post-orthonormalization value $6.661439\times10^{-15}$ was measured
over the complete normalized basis and therefore also bounds its occupied
block. These quantities are evaluated directly from the transported
orbitals before and after metric orthonormalization. The residual before orthonormalization
quantifies the finite PAW/plane-wave transfer representation; these tests
do not separate its individual approximation errors.

The 472-band transport used for the quoted secondary overlap
has corresponding occupied and post-orthonormalization residuals
$1.398739\times10^{-3}$ and $6.001563\times10^{-15}$.
These two numerical realizations of the same physical sudden reference
have squared determinant overlap 0.999987902. Using the 700-band sudden reference for both screening states
gives $P_1=0.629934046$ and $P_2=0.295789973$; the spectra use the
secondary weight 0.295338728 from the 472-band construction.
The local-density and seven-pair analyses
below use the common 700-band reference.

\subsection{\texorpdfstring{CeO$_2$}{CeO2}: two stationary screening states}
The primary charged state is obtained by ordinary energy-ordered occupation of the self-consistent KS orbitals. The secondary is located with the initial maximum overlap method (IMOM), which selects occupied orbitals by their overlap with a fixed reference rather than by eigenvalue order alone~\cite{Gilbert,Barca}. The reference used to select orbitals is held fixed; the electronic density and Hubbard occupation matrix are updated during the optimization.
In the final KS--IMOM stage for the CeO$_2$ secondary state, the Hubbard
matrix is updated every 20 iterations with unit mixing coefficient,
replacing it by the newly calculated matrix at each update.
The charge criterion is reached after 60 iterations, with an SCF residual
estimator of $9.2\times10^{-7}$~Ry.

For each spin, $M_{j\sigma}=C_{j\sigma}^{\dagger}S_{Z+1}C_{\mathrm{sud},\sigma}$. The singular values $s_{ji\sigma}$ of this matrix determine
\begin{equation}
 P_j=|\langle\Phi_j|\Phi_{\mathrm{sud}}\rangle|^2
 =\prod_{i\sigma}s_{ji\sigma}^{2}.
\end{equation}
All overlaps in Table~\ref{tab:overlaps} compare normalized determinants in the same PAW representation. The secondary charged-state energy is 11.7275~eV above the primary.
\begin{table}[H]
\centering\small
\caption{CeO$_2$ pairwise determinant overlaps used for the plotted weights.
The primary/sudden row uses the 700-band transport and the secondary/sudden
row uses the 472-band transport, whose difference is quantified above.
The first two squared overlaps supply approximate sudden weights in the
independent-channel model. The third measures the residual
nonorthogonality between screening states.}\label{tab:overlaps}
\begin{tabular}{lrrrr}\toprule
Comparison & Amplitude & Squared overlap & $\min s_i^\uparrow$ & $\min s_i^\downarrow$\\\midrule
Primary / sudden & 0.793684 & 0.629934 & 0.948650 & 0.928840\\
Secondary / sudden & 0.543451 & 0.295339 & 0.912986 & 0.912916\\
Secondary / primary & 0.270623 & 0.073237 & 0.750469 & 0.714254\\\bottomrule
\end{tabular}
\end{table}
These are pairwise squared overlaps, not probabilities of mutually orthogonal outcomes. In particular, the remainder of their sum from unity does not determine the weight of a missing screening state. Their ratio is $P_2/P_1=0.46884$; the corresponding spectra retain their individual transition strengths before they are combined.

\subsection{\texorpdfstring{CePO$_4$}{CePO4}: stationary-state search and local Hubbard checks}
\label{sec:cepo4-ns-consistency}
An initial search using staged SCF and fixed-reference KS--IMOM
preparations was followed by two ordinary SCF branches with Gaussian
occupations and freely updated Hubbard matrices. These converged to
the primary and alternative secondary magnetic solutions.
The magnetic neutral-$Z$ state supplies the occupied subspace transported
without orbital relaxation into the $Z+1$ PAW representation. Guided SCF
and fixed-reference KS--IMOM preparations provide the intermediate
densities and local matrices used to initialize the final optimizations.

The occupation-guided preparation starts from orbitals calculated with
Gaussian smearing of 0.05~Ry. To construct the Hubbard guide, bands
1--81 and 83 in spin up and 1--82 in spin down are assigned unit
occupation, with the remaining bands empty. This integer filling is
used to prepare the local matrix, not imposed on the subsequent SCF
charge density, whose occupations remain Gaussian. The guide steers
the search rather than prescribing the final fractional occupations.
Optimization with freely mixed Hubbard matrices retains a primary basin
with one nearly filled absorber $4f$ direction and additional fractional
occupation. The alternative is initialized using periodic Hubbard updates
that mix 10\% of the newly calculated matrix with 90\% of the current
matrix; the final guided stage has an interval of 22 iterations.
Releasing the matrix gives a solution with a second nearly filled
absorber $4f$ direction. Neither a periodic matrix hold nor MOM orbital
selection is used in the final SCF optimization of either state.

Both final calculations use Gaussian smearing of 0.02~Ry, contain
164 electrons and 200 bands per spin, use the shifted $3\times3\times3$
mesh, and constrain total magnetization
to zero. Local Thomas--Fermi preconditioning and density mixing
coefficient 0.04 are used, with four history vectors for the primary
and six for the alternative. Davidson starts at $10^{-8}$~Ry with
full band accuracy and subspace factor two. The convergence criteria and
diagnostics of the resulting states are listed in Table~\ref{tab:cepo4-stationary}.

\begin{table}[H]\centering\small
\caption{Final charged CePO$_4$ references. Both use freely updated
Hubbard matrices. Energies are QE's $F=E-TS$ estimates with the same
Gaussian smearing of 0.02~Ry.
The iteration counts refer only to the final freely updated SCF stages,
not the entire search. The listed local eigenvalues are the three largest
in the absorber's spin-up $4f$ block.}\label{tab:cepo4-stationary}
\begin{tabular}{lrr}\toprule
Quantity & Primary & Alternative\\\midrule
SCF threshold (Ry) & $1.4\times10^{-4}$ & $4.0\times10^{-5}$\\
Final SCF estimator (Ry) & $1.1305\times10^{-4}$ & $3.980\times10^{-5}$\\
Iterations in final SCF stage & 4 & 430\\
$F$ (Ry) & $-2846.61653523$ & $-2846.65921008$\\
Absorber $n_{4f}$ & 1.783270 & 2.010255\\
Absorber $4f$ moment ($\mu_B$) & 1.593755 & 1.906052\\
$\lambda_1$ & 0.997762 & 0.990449\\
$\lambda_2$ & 0.320842 & 0.920987\\
$\lambda_3$ & 0.316142 & 0.015156\\
\bottomrule\end{tabular}\end{table}

The alternative is lower by 0.04267485~Ry, or 0.58062~eV.
The primary label therefore refers to its stronger relation to the
sudden reference, not the total-energy ordering. The additional nearly
filled local direction in the alternative describes stronger absorber
localization; the projected populations are not assigned directly to a
new local spectroscopic sector.

For the same globally orthonormalized PAW atomic basis, the local
matrix is reconstructed from the saved KS orbitals and their occupations.
Let $|p_m^I(\bm k)\rangle$ denote the Bloch sum of a local atomic $4f$
function on Ce site $I$, after L\"owdin orthonormalization in the PAW
metric. The index $m$ runs over its seven angular components. These
local orbitals are distinct from the beta projectors supplied by the
PAW input. Their projection amplitudes are
$a_{m,b\bm k\sigma}^{I}=\langle p_m^I(\bm k)|S(\bm k)|\psi_{b\bm k\sigma}\rangle$,
where $b$ labels a KS band and $\sigma$ its spin. With occupations
$f_{b\bm k\sigma}$ per spin orbital and normalized k-point weights
$w_{\bm k}$, the local matrix is
\begin{equation}
 D_{mm'}^{I\sigma}=
 \sum_{\bm k}w_{\bm k}\sum_b f_{b\bm k\sigma}
 a_{m,b\bm k\sigma}^{I}a_{m',b\bm k\sigma}^{I*},
 \qquad \sum_{\bm k}w_{\bm k}=1.
 \label{eq:hubbard-density}
\end{equation}
Collinear QE uses the real symmetric part as the Hubbard matrix.
The trace of each spin block is its local projected $4f$ population;
the sum and difference of the two traces give the total population
and spin moment, respectively, with the latter reported in $\mu_B$.
The imaginary remainder is at most $2.39\times10^{-9}$ in these
$4f$ blocks. The reconstruction preserves all off-diagonal elements.
With actual Gaussian occupations, it agrees with the saved matrix
$n_{\rm save}^{I\sigma}$ below $2.7\times10^{-15}$ in Frobenius norm
per Ce, including both spins.

This agreement verifies the consistency of the collected wavefunctions,
occupations and saved spectral potential. At convergence, QE saves
the output density and output Hubbard matrix and rebuilds their
potential. Agreement with $n_{\rm save}$ is consequently not a separate
measurement of the input--output residual in the last Davidson step.
The iterative convergence is characterized by the SCF estimator in
Table~\ref{tab:cepo4-stationary}.

Two filling conventions are evaluated on the same final primary
wavefunctions. The Gaussian convention retains the fractional occupations
from the SCF at 0.02~Ry and reconstructs the spectral background.
The integer convention occupies the first 82 orbitals of each spin and
$\bm k$ with unit weight, characterizing the determinant used for overlaps.
It is a diagnostic and is not substituted into the spectral potential.
Define
\begin{align}
 r_I&=\left[\sum_\sigma
 \|\operatorname{Re}D^{I\sigma}-n_{\rm save}^{I\sigma}\|_F^2\right]^{1/2},\\
 \eta_I&=\frac{r_I}{[\sum_\sigma\|n_{\rm save}^{I\sigma}\|_F^2]^{1/2}} .
\end{align}
The integer changes mostly redistribute spin-up $4f$ occupation between
the absorber and its same-spin Ce neighbor at 4.149~\AA\ (Ce3), while
retaining their leading orbital orientations.
The two columns in Table~\ref{tab:cepo4-ns-consistency} therefore answer
different questions. The Gaussian column checks that the saved orbitals
and occupations reproduce the matrix used for spectroscopy. The integer
column measures what changes when fractional occupations are replaced by
a filled set of 82 orbitals per spin and k point, without changing the
orbitals themselves. Its larger values on the absorber and Ce3 are not
residuals of an unconverged SCF iteration: this integer filling was never
fed back into the spectral potential. On the absorber the change is mainly
in the two fractionally occupied directions; on Ce3 it mainly raises the
leading occupation, as shown in Table~\ref{tab:cepo4-ns-leading}.
The squared overlaps of the leading local directions remain above 0.99996.
Thus the orientations are stable, but the two fillings do not define
identical local populations. The integer representation is used only
for the stated determinant-overlap diagnostic, not as a replacement
for the Gaussian screening density.

\begin{table}[H]\centering\footnotesize
\caption{Reconstruction of the primary saved Hubbard matrix. G denotes
actual Gaussian occupations; int.\ denotes unit occupation of the first
82 orbitals in each spin and $\bm k$ block.
The relative norm is in percent. $\Delta N_I$ (electrons) and
$\Delta M_I$ ($\mu_B$) are computed as integer representation minus
saved matrix; they are not SCF residuals. $q_I$ is the squared overlap
between the leading majority-spin eigenvectors of the integer and
saved matrices. Gaussian differences are checks of
reconstruction consistency, not additional SCF convergence thresholds.
Ce3 is the same-spin neighbor 4.149~\AA\ from the absorber; the site
labels are defined in Sec.~\ref{sec:settings}.}
\label{tab:cepo4-ns-consistency}
\begin{tabular}{lrrrrrr}\toprule
Site & $r_I$ (G) & $r_I$ (int.) & $100\eta_I$ (int.) &
$\Delta N_I$ & $\Delta M_I$ & $q_I$\\\midrule
Absorber & $1.97\times10^{-15}$ & 0.219566 & 20.04 & $-0.309900$ & $-0.309900$ & 0.99999987\\
Ce2 & $1.39\times10^{-15}$ & 0.000083 & 0.0083 & $+0.000095$ & $+0.000095$ & 1.00000000\\
Ce3 & $2.64\times10^{-15}$ & 0.250885 & 33.74 & $+0.250985$ & $+0.250985$ & 0.99996149\\
Ce4 & $1.18\times10^{-15}$ & 0.000075 & 0.0076 & $+0.000083$ & $+0.000083$ & 1.00000000\\
\bottomrule\end{tabular}\end{table}

\begin{table}[H]\centering\footnotesize
\caption{Leading majority-spin local $4f$ eigenvalues of the primary
reference, evaluated with the SCF Gaussian occupations and with the
integer occupations used for the determinant-overlap diagnostic.}\label{tab:cepo4-ns-leading}
\begin{tabular}{llrrrrrr}\toprule
 & & \multicolumn{3}{c}{Gaussian} & \multicolumn{3}{c}{Integer}\\
Site & Spin & $\lambda_1$ & $\lambda_2$ & $\lambda_3$ &
$\lambda_1$ & $\lambda_2$ & $\lambda_3$\\\midrule
Absorber & $\uparrow$ & 0.997762 & 0.320842 & 0.316142 & 0.997830 & 0.173746 & 0.155473\\
Ce2 & $\downarrow$ & 0.996916 & 0.014462 & 0.013279 & 0.996916 & 0.014462 & 0.013279\\
Ce3 & $\uparrow$ & 0.741975 & 0.020955 & 0.019603 & 0.992750 & 0.021004 & 0.019618\\
Ce4 & $\downarrow$ & 0.996975 & 0.014240 & 0.013407 & 0.996975 & 0.014240 & 0.013407\\
\bottomrule\end{tabular}\end{table}

Over the four complete iterations of the final primary restart, the
total $4f$ trace spans are 0.00056, 0.00001, 0.00138 and 0.00001 on
the absorber, Ce2, Ce3 and Ce4. The largest resolved change of a printed
local eigenvalue is 0.001, on Ce3. Traces and eigenvalues are printed to
$10^{-5}$ and $10^{-3}$, respectively. The reported total energies span
$4.2569\times10^{-4}$~Ry, or 5.79~meV for the 24-atom cell.
This stability over the final SCF iterations and the SCF estimator describe different
numerical quantities.

For the 27 equally weighted shifted mesh points, define
$p_{\bm k}=\prod_\sigma|\det M_{\bm k\sigma}|^2$ using 82 occupied
orbitals per spin. The common mesh diagnostic is
\begin{equation}
 P_{\mathrm{geom}}=
 \exp\left[\frac{1}{27}\sum_{\bm k}\log p_{\bm k}\right].
\end{equation}
All three comparisons use the same PAW metric and physical $\bm k$.
The singular values reproduce the determinant products independently.

\begin{table}[H]\centering\small
\caption{CePO$_4$ integer-determinant overlap diagnostics. The spin
factors are geometric means; their product gives $P_{\rm geom}$.}
\label{tab:cepo4-overlaps}
\begin{tabular}{lrrrr}\toprule
Pair & $P_{\rm geom}$ & Spin up & Spin down & Per-$\bm k$ range\\\midrule
Primary / sudden & 0.722581 & 0.758218 & 0.953000 & 0.693773--0.798256\\
Alternative / sudden & 0.001376 & 0.001443 & 0.953528 & 0.000042--0.009773\\
Alternative / primary & 0.005979 & 0.006063 & 0.986151 & 0.000246--0.032055\\
\bottomrule\end{tabular}\end{table}

The loss of alternative/sudden overlap is predominantly spin up,
consistent with its changed local majority-spin occupation. The two
stationary references are distinct, but only the primary retains a
large sudden overlap under this convention. No separate spectrum or
injected-state calculation is assigned to the alternative here.
The full-mesh log products are $-8.77298$, $-177.88728$ and $-138.22474$
in the table's order. Neither the geometric mean nor the full product
is an exclusive absorption probability for the fractional Gaussian
density; these numbers do not rescale the CePO$_4$ spectra.
The search is not claimed to exhaust all stationary screening states.

\FloatBarrier
\section{Local orbital character of the screening response}
\label{sec:local}
The local spectroscopic sector is not assigned from the total projected
$4f$ population alone. All states are compared using the same local
atomic orbitals $p_m^I(\bm k)$ defined before
Eq.~\eqref{eq:hubbard-density}. For screening state $j$, the spin-resolved
local density matrix can be written explicitly as
\begin{equation}
 \begin{aligned}
 D^{(j\sigma)}_{mm'}={}&\sum_{\bm k}w_{\bm k}\sum_n f^{(j)}_{n\bm k\sigma}
 \langle p_m(\bm k)|S(\bm k)|\psi^{(j)}_{n\bm k\sigma}\rangle\\
 &\times\langle\psi^{(j)}_{n\bm k\sigma}|S(\bm k)|p_{m'}(\bm k)\rangle,
 \end{aligned}
 \label{eq:local-density}
\end{equation}
where the site index $I$ is suppressed and $n$ labels the KS bands.
The factors of $S$ express the PAW scalar product; they do not introduce
another projection operation. The trace of $D$ gives the local projected
population. For an integer-occupied determinant, the largest eigenvalue
also bounds the $4f$ weight of a normalized linear combination within the
occupied subspace. This distinguishes distributed covalent $4f$ character
from an almost completely occupied localized orbital.

\begin{table}[H]
\centering\small
\caption{Local absorber populations and largest $4f$ density-matrix eigenvalues reported in the main article and the local-orbital analysis. Eigenvalue pairs are ordered spin up, then spin down.}\label{tab:populations}
\begin{tabular}{p{0.34\linewidth}rrp{0.26\linewidth}}\toprule
State & $n_{4f}$ & $n_{5d}$ & Largest $4f$ eigenvalues\\\midrule
CeO$_2$ sudden reference & 0.652737 & 1.806993 & 0.133387 and 0.133443\\
CeO$_2$ primary & 1.735684 & 1.804062 & 0.394 and 0.447\\
CeO$_2$ secondary & 0.067359 & 2.894459 & About 0.017 in each spin\\
CePO$_4$ primary & 1.783270 & 1.416713 & 0.997762 and 0.023821\\\bottomrule
\end{tabular}
\end{table}
In \ceo, the increase from the sudden reference to the primary state is about 1.083 in projected $4f$ population, but no local eigenvalue approaches unity. Seven corresponding orbital pairs account for 92.3\% of the net increase. The two largest contributions raise their $4f$ weights from about 0.11 to about 0.37 and 0.42 while retaining O $2p$ character. Relative to the primary state, the secondary loses 1.668325 of absorber $4f$ weight, gains 1.090396 of absorber $5d$ weight, and gains 0.490239 in the sum of O $2p$ projections. These quantities describe projected orbital character rather than a unique atomic charge partition.
The maximum primary eigenvalue 0.394157 belongs to spin up and 0.447157
to spin down. All CeO$_2$ states use the same set of 896 globally
L\"owdin-orthonormalized PAW atomic orbitals. The CePO$_4$ values use its
own globally orthonormalized projection basis and the saved fractional
occupations; they describe the output density checked against the saved
spectral Hubbard matrix in Sec.~\ref{sec:cepo4-ns-consistency}.
\begin{table}[H]\centering\small
\setlength{\tabcolsep}{3pt}
\caption{Complete local $4f$ eigenvalue spectra, descending within each spin.
Ce1 is the absorber in CePO$_4$.}\label{tab:local-eigenvalues}
\begin{tabular}{llrrrrrrr}\toprule
State/site & Spin & $\lambda_1$ & $\lambda_2$ & $\lambda_3$ & $\lambda_4$ & $\lambda_5$ & $\lambda_6$ & $\lambda_7$\\\midrule
\multicolumn{9}{l}{(a) CeO$_2$}\\
Sudden & $\uparrow$ & 0.133387 & 0.044145 & 0.044101 & 0.044099 & 0.020203 & 0.020202 & 0.020202\\
Sudden & $\downarrow$ & 0.133443 & 0.044146 & 0.044102 & 0.044101 & 0.020203 & 0.020202 & 0.020201\\
Primary & $\uparrow$ & 0.394157 & 0.198666 & 0.194797 & 0.101458 & 0.032106 & 0.031982 & 0.031338\\
Primary & $\downarrow$ & 0.447157 & 0.080861 & 0.067844 & 0.066594 & 0.029666 & 0.029620 & 0.029440\\
Secondary & $\uparrow$ & 0.016803 & 0.004998 & 0.004994 & 0.004994 & 0.000629 & 0.000629 & 0.000629\\
Secondary & $\downarrow$ & 0.016804 & 0.004999 & 0.004995 & 0.004995 & 0.000629 & 0.000629 & 0.000629\\
\midrule
\multicolumn{9}{l}{(b) CePO$_4$}\\
Ce 1 & $\uparrow$ & 0.997762 & 0.320842 & 0.316142 & 0.020370 & 0.014681 & 0.011705 & 0.007011\\
Ce 1 & $\downarrow$ & 0.023821 & 0.022162 & 0.014392 & 0.013389 & 0.008575 & 0.007000 & 0.005418\\
Ce 2 & $\uparrow$ & 0.013457 & 0.012513 & 0.009266 & 0.007799 & 0.005205 & 0.004101 & 0.003617\\
Ce 2 & $\downarrow$ & 0.996916 & 0.014462 & 0.013279 & 0.008363 & 0.006931 & 0.004796 & 0.003799\\
Ce 3 & $\uparrow$ & 0.741975 & 0.020955 & 0.019603 & 0.014013 & 0.011692 & 0.006660 & 0.005446\\
Ce 3 & $\downarrow$ & 0.020046 & 0.018377 & 0.013892 & 0.011057 & 0.006735 & 0.006007 & 0.004911\\
Ce 4 & $\uparrow$ & 0.013326 & 0.012524 & 0.009340 & 0.007830 & 0.005265 & 0.004120 & 0.003607\\
Ce 4 & $\downarrow$ & 0.996975 & 0.014240 & 0.013407 & 0.008411 & 0.006904 & 0.004850 & 0.003764\\
\bottomrule\end{tabular}\end{table}
The total O $2p$ weights for the sudden, primary and secondary CeO$_2$
states are 304.868703, 304.559634 and 305.049873; the corresponding summed
nonabsorbing-Ce $4f$ weights are 21.825739, 21.660406 and 21.693489.
\begin{table}[H]\centering\footnotesize
\caption{Seven corresponding orbital pairs with the largest absolute
absorber $4f$ changes for the primary state. Pair indices follow descending
singular values within each spin. Oxygen weights sum all O $2p$ projections.}
\label{tab:seven-pairs}
\begin{tabular}{lrrrrrr}\toprule
Spin, pair & $s_i$ & $f_{\rm sud}$ & $f_1$ & $\Delta f$ & O $2p_{\rm sud}$ & O $2p_1$\\\midrule
$\downarrow$, 384 & 0.928840 & 0.109341 & 0.423166 & 0.313825 & 0.767824 & 0.504355\\
$\uparrow$, 384 & 0.948650 & 0.111420 & 0.371700 & 0.260280 & 0.767928 & 0.550471\\
$\uparrow$, 383 & 0.965021 & 0.036857 & 0.192057 & 0.155200 & 0.855828 & 0.727724\\
$\uparrow$, 382 & 0.966200 & 0.035632 & 0.186604 & 0.150973 & 0.857123 & 0.732834\\
$\uparrow$, 381 & 0.992581 & 0.037953 & 0.095540 & 0.057587 & 0.849817 & 0.809214\\
$\downarrow$, 383 & 0.996317 & 0.038189 & 0.075312 & 0.037124 & 0.849538 & 0.827232\\
$\downarrow$, 382 & 0.998064 & 0.038600 & 0.062842 & 0.024242 & 0.840515 & 0.830219\\
\bottomrule\end{tabular}\end{table}
The seven increments sum to 0.999230737, or 92.2696\% of the net increase
1.082947103. All remaining pairs contribute 0.083716367. The positive
increments over all pairs sum to 1.085533790 and the negative increments
to $-0.002586687$; the percentage uses their signed net sum.
There are four up-spin and three down-spin pairs, each in a nondegenerate
SVD block under the $10^{-8}$ singular-value grouping tolerance.
Seven therefore does not denote the seven angular functions of one spin
shell.

For the determinant overlap, the additive loss associated with each corresponding occupied pair is $w_{ji\sigma}=-\log(s_{ji\sigma}^2)$. Sorting these losses over both spins gives the cumulative fraction $F_r=\sum_{a=1}^{r}w_{(a)}/[-\log P_j]$.
\begin{table}[H]
\centering\small
\caption{Concentration of overlap loss over the 768 occupied principal directions. $N_q$ is the number required to account for $q$ percent of $-\log P_j$.}\label{tab:loss}
\begin{tabular}{lrrrrrr}\toprule
State & $-\log P_j$ & $F_2$ & $F_5$ & $N_{50}$ & $N_{90}$ & $N_{99}$\\\midrule
Primary & 0.462140 & 0.5476 & 0.8827 & 2 & 7 & 71\\
Secondary & 1.219632 & 0.2987 & 0.4427 & 7 & 24 & 48\\\bottomrule
\end{tabular}
\end{table}
\begin{figure}[H]\centering
\includegraphics[width=0.92\linewidth]{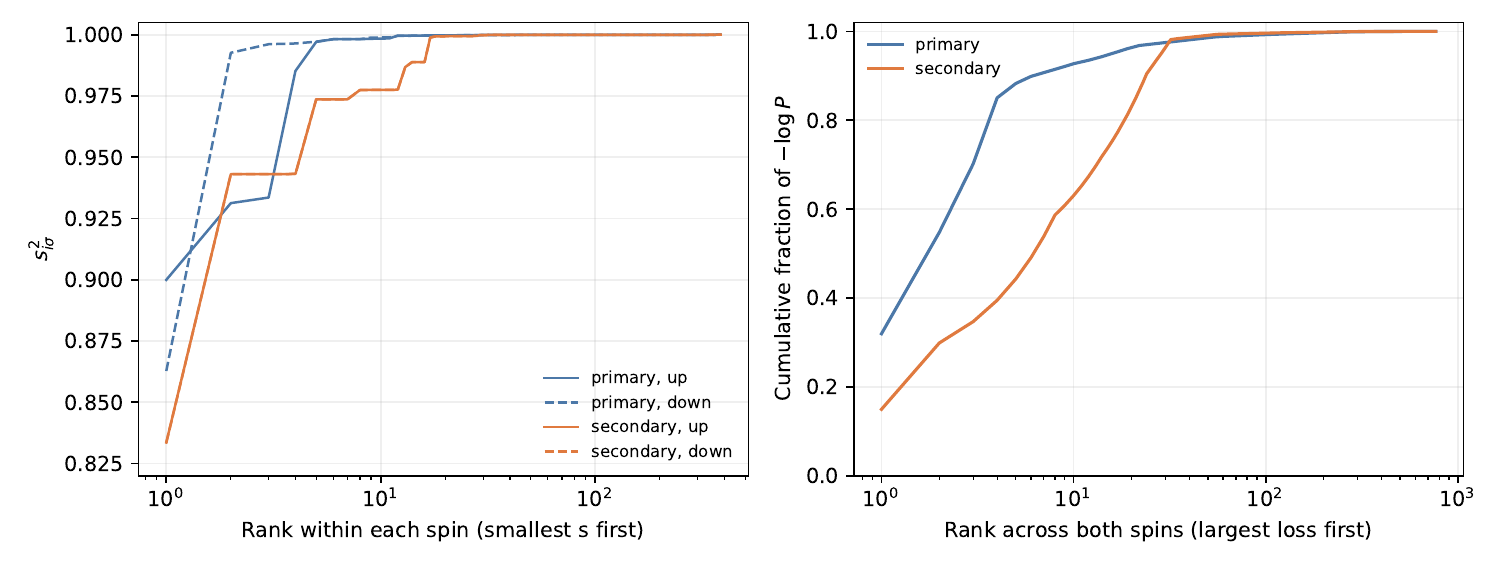}
\caption{Squared singular values of the occupied-space overlaps (left) and cumulative overlap loss (right). The primary loss is concentrated in fewer principal directions than the secondary loss. A principal direction can itself combine many KS orbitals; the curves are not a real-space charge decomposition.}\label{fig:loss}
\end{figure}

\FloatBarrier
\section{Resonance wavepackets and neutral-state relaxation}
\label{sec:relax}
The neutral-state optimizations in this section concern CeO$_2$ and are
separate from the fixed-background spectral calculations. We begin with
the primary charged screening determinant, containing 384 occupied
orbitals in each spin channel. To construct a neutral trial state, we
select one additional orbital from either the $4f$ resonance or the
white-line resonance of that background.

The excitation source is filtered around the selected energy by a finite
Chebyshev expansion of the spectral Hamiltonian, with Jackson
damping~\cite{Weisse}. This produces a wavepacket with finite energy
resolution, not an individual exact eigenstate. We retain its photoelectron
component in the dominant local channel, remove its overlap with the
already occupied KS orbitals using the PAW scalar product, and normalize
what remains. Both selected orbitals have spin down. Occupying the added
orbital with one electron, together with the original occupied orbitals,
defines a 769-electron neutral Slater determinant: 384 spin-up and 385
spin-down orbitals. The charge density, PAW augmentation and Hubbard
matrix are reconstructed from this occupied set before relaxation starts.

Each neutral determinant is then optimized independently with KS--IMOM.
At every SCF iteration the current Hamiltonian is diagonalized and the
occupied subset is selected by its overlap with that run's fixed initial
neutral reference, rather than solely by eigenvalue order. The electron
counts and integer occupations are retained, but the selected orbitals
themselves can change. The density and the periodically updated Hubbard
matrix therefore relax even though the orbital-selection reference stays
fixed. This differs both from propagating a spectrum in a fixed potential
and from freezing the occupied wavefunctions. The fractional fillings
used for the CePO$_4$ screening density, and their separate integer
representation for overlaps, were described in
Sec.~\ref{sec:cepo4-ns-consistency}; no analogous pair of neutral-state
relaxations is used for CePO$_4$ here.
Both injected CeO$_2$ packets use 10000 moments over $[-63.5,165]$~Ry
on the primary $\Gamma$ XSpectruplet Hamiltonian's internal scale.
Their filter centers are $-20.47746847$~eV for $4f$ and
$-12.22156369$~eV for the white line on that scale, before subtraction
of the white-line energy.
The Jackson delta-kernel amplitude has FWHM about 1.033 and 1.036~eV
at those energies. This specifies the filter kernel, not the width of
the packet's squared energy distribution after multiplication by the source
spectral weight. No independent Gaussian target filter is used.
At each energy the dominant local channel is selected by its largest
PAW norm.
Both packets are constructed at the electronic $\Gamma$ point.
The $4f$ selection uses photon direction
$\widehat{\bm q}=(4,3,2)/\sqrt{29}$ and
polarization $\boldsymbol\epsilon=(1,-2,1)/\sqrt6$ in the cell Cartesian
frame. The white-line packet instead uses
$\widehat{\bm q}=\hat{\bm z}$ and
$\boldsymbol\epsilon=\hat{\bm x}$, as do the main-figure spectra.
These specify the packet initialization geometries; no equivalence
between the two orientations is assumed.

During these neutral-state optimizations, the density-mixing coefficient
is 0.02. The Hubbard matrix is held fixed within each 20-iteration block
and then mixed equally with the newly reconstructed matrix; at least
seven such updates are required. The diffuse white-line packet requires
a larger unoccupied manifold: the reported optimization uses 1000 KS
bands, whereas a 472-band space did not adequately represent that packet.

\begin{table}[H]
\centering\small
\caption{CeO$_2$ relaxed neutral-state energies used in the gap comparison.
The entries in Ry are QE total cell energies with 769 electrons and
the same PAW inputs; they are not spectral pole or photon energies.
The fixed-background spectral separation is evaluated at $\Gamma$.}\label{tab:energies}
\begin{tabular}{lr}\toprule
Quantity & Value\\\midrule
Relaxed $4f$ plateau & $-18128.99150606$~Ry\\
$4f$ state after an additional Hubbard-matrix update & $-18128.99157934$~Ry\\
Relaxed white-line state & $-18128.21938809$~Ry\\
White-line--$4f$ separation, plateau & 10.5052002~eV\\
White-line--$4f$ separation, after the additional update & 10.5061972~eV\\
Fixed-background spectral separation & 8.255905~eV\\\bottomrule
\end{tabular}
\end{table}
For the \ceo\ $4f$ calculation, the SCF residual estimator reported by QE is $1.314\times10^{-5}$~Ry at the selected energy plateau. A further Hubbard-matrix update changes its total energy by $7.33\times10^{-5}$~Ry, approximately 1~meV. The white-line optimization completes 155 iterations and seven Hubbard updates with an SCF residual estimator of $1.833\times10^{-5}$~Ry. Its squared determinant overlap with the initial neutral reference is 0.405117. These quantities characterize the numerical state selection and energy stability; the fixed-reference orbital criterion does not freeze the optimized density or Hubbard matrix.

The relaxed total-energy difference is compared with the 8.26-eV separation
of the fixed-background spectral maxima. Their difference, about 2.25~eV,
defines the estimated gap correction. Both neutral endpoints are optimized,
so the white-line relaxation is included in this comparison. The estimate
combines total-energy and spectral descriptions; its transfer to CePO$_4$
is the heuristic assumption described in Sec.~\ref{sec:centroid}.
Both final-state runs use integer input occupations, so no smearing-entropy
term is present in this energy difference. With
$1~{\rm Ry}=13.605693123$~eV, the two listed plateau energies give
10.5052002~eV. Using the $4f$ energy after the additional Hubbard-matrix update
gives 10.5061972~eV. The 10.505--10.506-eV range describes the
approximately 1-meV sensitivity to that update, without estimating the
overall physical uncertainty. Both values round to 10.51~eV in the main text.
Both runs request a charge threshold $2\times10^{-5}$~Ry and
an initial Davidson eigenvalue threshold $10^{-8}$~Ry.
The SCF residual estimator reported by QE is obtained from the charge-mixing
calculation; it is not a rigorous bound on the total-energy error
or the norm of every eigenvector residual. The $4f$ endpoint is the stable
pre-update plateau; the white-line run satisfies its stopping criterion
after the minimum number of Hubbard updates.

\FloatBarrier
\section{Radial-source and core spin--orbit tests}
\label{sec:source}
For each final angular momentum, the transition source is expanded in
all-electron radial eigenfunctions inside a spherical box. The cutoff
specifies which box eigenstates enter this expansion, not photon energies
or offsets from the spectral white line. To localize the source around
the absorber, its radial function is left unchanged up to 1.9~bohr,
smoothly reduced to zero between 1.9 and 2.85~bohr, and set to zero beyond
2.85~bohr. This spatial cutoff is applied to the all-electron source,
before conversion to the PAW representation.

As explained in Appendix~\ref{main-app:paw-source} of the main article,
applying the fixed PAW transformation $\mathcal T^\dagger$ and expanding
in plane waves gives the transition amplitudes $d$. The initial
plane-wave state $\widetilde b$ is obtained by solving
$S_{\rm PW}\widetilde b=d$, so that its PAW overlaps reproduce those
of the all-electron source. Thus the original PAW partial waves and projectors
define the mapping at every radial cutoff; no independently pseudized
box functions replace that mapping. The relative residual of this linear solve
is below $10^{-10}$ for all reported spectra. The main spectra use a
20-Ry radial-energy cutoff.

Here $K$ is the excitation multipole rank and $L$ the photoelectron
angular momentum at the absorber. For the retained dipole source
($L=2$, $K=1$), the 20-Ry basis contains seven states with node counts
2 through 8 and energies
\begin{equation}
 (-0.1773,\;0.9592,\;2.8583,\;5.5149,\;8.8895,\;12.9537,\;17.6896)~\mathrm{Ry}.
\end{equation}
The stated local pseudization radii are 1.5~bohr for $d$ and 0.9~bohr for $f$, with a common augmentation radius of 1.9~bohr. The radial-node analysis places the first additional internal nodes at 12.953750~Ry for $d$ and 10.394483~Ry for $f$. These values are below the adopted cutoff and do not establish a node-free range extending to 20~Ry.
The radial eigenfunctions are regular at the origin and vanish at
$R_{\rm box}=5.7400$~bohr, approximately three times the augmentation
radius. They are normalized inside this box; the source-expansion
integrals extend to 5.7~bohr. The finite box supplies a discrete basis
for the radial source, not a confining boundary for the propagated
photoelectron in the crystal.

The quadrupole basis has node counts 0--7 and energies
\begin{equation}
 \begin{split}
 (&-0.764812,\ 0.855040,\ 2.224893,\ 4.275072,\\
  &7.006553,\ 10.394483,\ 14.422263,\ 19.080439)\ {\rm Ry}.
 \end{split}
\end{equation}
The atomic pseudization radii 0.9~bohr ($f$) and 1.5~bohr ($d$)
remain distinct from the 1.9-bohr augmentation radius. These describe
the atomic construction; the PAW transformation used for the sources
is held fixed throughout the cutoff comparison.

The core $2p$ spin--orbit scale of 0.2 used in the main calculations is
chosen for numerical efficiency in the $L_3$ region, not by fitting the
experimental spectrum. A stronger core splitting increases the eigenvalue
range to be represented by polynomial recursions; in a Chebyshev expansion,
more moments are then needed to retain a given energy resolution.
The tests below assess the sensitivity of the $L_3$ response to this
numerical compromise.

The five sensitivity controls use the coherent E1+E2 excitation source
on the same primary charged CeO$_2$ reference at $\Gamma$.
All use 1000 Lanczos iterations and Lorentzian HWHM 0.3~eV.
The spectra are reconstructed from full-precision recurrence coefficients
using the same procedure for every parameter choice. Each curve is aligned
and normalized at its own white-line maximum, after restoring source
strength. The source mapping is the same fixed-PAW construction as in
the main calculations. The relative E1/E2 normalization is fixed by
the transition operators, not fitted to experiment.

\begin{table}[H]\centering\small
\caption{E1+E2 radial-cutoff sensitivity at core-spin--orbit scale 0.2.
$N_d$ is the retained dipole radial-basis size, $r$ the source-norm
amplitude ratio relative to 20~Ry, and $r^2$ the source-strength ratio.
$D$ compares white-line-normalized L1000 profiles with the 20-Ry reference;
$D_{500/1000}$ compares recurrence lengths. Both are relative $L^2$
differences, expressed as percentages over $[-18,30]$~eV relative
to each white line.}\label{tab:radial}
\begin{tabular}{rrrrrr}\toprule
Cutoff (Ry) & $N_d$ & $r$ & $r^2$ & $D$ (\%) & $D_{500/1000}$ (\%)\\\midrule
10 & 5 & 0.732068 & 0.535924 & 2.668 & 12.24\\
15 & 6 & 0.872539 & 0.761325 & 4.713 & 12.73\\
20 & 7 & 1.000000 & 1.000000 & 0.000 & 12.43\\
\bottomrule\end{tabular}\end{table}
The source strength retains a cutoff dependence, while the E2-dominated
pre-edge remains present at all three cutoffs. Its E1+E2 local maximum
lies between $-8.10$ and $-8.09$~eV relative to each white line.
The corresponding peak-to-white-line ratios are 0.0434, 0.0505 and
0.0539 at 10, 15 and 20~Ry. The unnormalized pre-edge peak decreases
by about 6.2\% from 10 to 20~Ry, whereas the white-line maximum decreases
by about 24.5\%. The increase of their ratio therefore does not imply
a growing absolute quadrupole peak. Between 15 and 20~Ry the unnormalized
pre-edge maximum changes by less than 1\%. These are E1+E2 peak values,
including the dipole background, not isolated E2 intensities.

\begin{table}[H]\centering\small
\caption{E1+E2 core-spin--orbit sensitivity at 20-Ry radial cutoff,
1000 Lanczos iterations and 0.3-eV HWHM. The relative $L^2$ differences
use the same window and conventions as Table~\ref{tab:radial}, with
scale 0.2 as reference for $D$. The last column gives the position of
the local E1+E2 pre-edge maximum relative to the white-line maximum
of the same spectrum. It is the largest interior local maximum in
$[-9,-7]$~eV, sampled on the existing 0.01-eV grid without peak fitting
or sub-grid interpolation.}\label{tab:so}
\begin{tabular}{rrrr}\toprule
Scale & $D$ (\%) & $D_{500/1000}$ (\%) &
$E_{\mathrm{pre}}^{\mathrm{E1+E2}}-E_{\mathrm{WL}}$ (eV)\\\midrule
0.1 & 0.778 & 13.27 & $-8.10$\\
0.2 & 0.000 & 12.43 & $-8.10$\\
1.0 & 3.638 & 18.00 & $-8.07$\\
\bottomrule\end{tabular}\end{table}
Across the tested core-spin--orbit scales, the E1+E2 pre-edge maximum
lies between $-8.10$ and $-8.07$~eV relative to its own white line,
spanning 0.03~eV.
The 500/1000-step differences are 12.24--18.00\% over the full window,
but 1.76--2.81\% in the pre-edge window $[-12,-5]$~eV. The parameter
sensitivities are comparisons at a common finite recurrence length,
not a certification of fully converged line shapes or absolute intensities.

\begin{figure}[H]\centering
\includegraphics[width=0.94\linewidth]{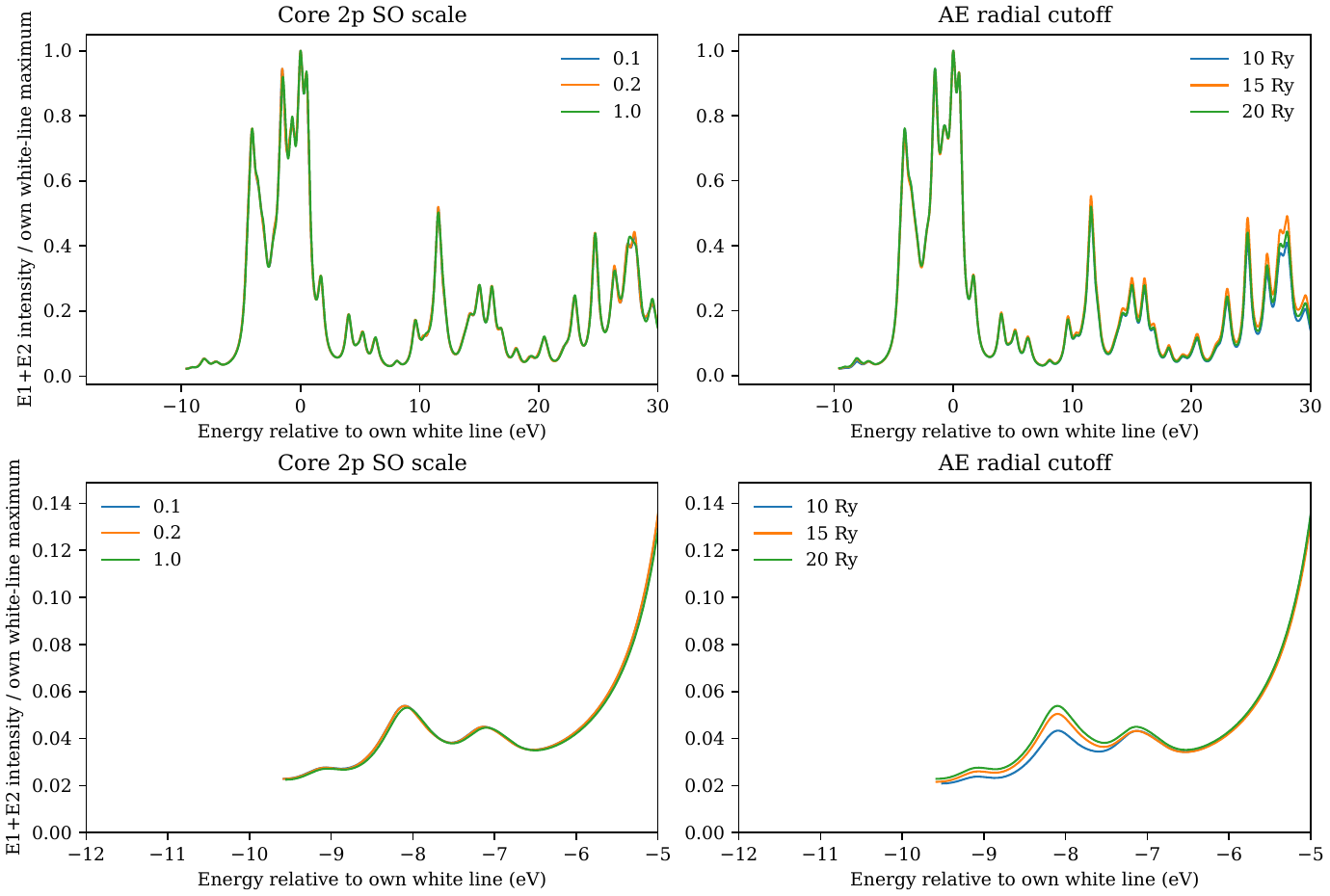}
\caption{E1+E2 sensitivity to the core-spin--orbit scale (left) and
radial cutoff (right) for the primary CeO$_2$ background at $\Gamma$.
Top: overview; bottom: pre-edge. All curves use L1000 and HWHM 0.3~eV,
with their own white-line maximum setting energy zero and unit intensity.
Display cutoffs at minima below the $4f$ feature lie between $-9.58$
and $-9.51$~eV. No cut is applied to the recurrence or to the numerical
comparisons. The PAW transformation is fixed in all cases.}\label{fig:sensitivity}
\end{figure}

For the two sensitivity tables, $r=\|b\|_S/\|b_{\rm ref}\|_S$ and
\begin{equation}
 D=\frac{\|y-y_{\rm ref}\|_2}{\|y_{\rm ref}\|_2}.
\end{equation}
The uniform 0.01-eV grid covers $[-18,30]$~eV relative to each white line.
For $D_{500/1000}$, the denominator is the L1000 profile and both truncations
use that run's L1000 energy origin and normalization. These global
relative differences are not pointwise relative errors. Independent
white-line alignment tests line shapes, not absolute core-energy shifts.

\FloatBarrier
\section{Spectral resolution, averaging, and experimental reference}
\label{sec:spectral}
The spectral calculation acts repeatedly with the fixed excitation
Hamiltonian on the initial source. It does not re-optimize the occupied
KS orbitals or the Hubbard matrix. Lanczos represents this action by a
finite tridiagonal matrix; Chebyshev uses a sequence of polynomials of
the Hamiltonian. For a rescaled Hamiltonian $\widetilde H$ with spectrum
in $[-1,1]$, the latter sequence gives
$\mu_n=\langle b|T_n(\widetilde H)|b\rangle$, where $b$ is the initial
excitation source and $T_n$ the Chebyshev polynomial of order $n$.
Jackson damping suppresses oscillations caused by truncating this
sequence, at the cost of finite spectral resolution~\cite{Weisse}.
The 10000-moment CeO$_2$ calculation has an intrinsic full width at
half maximum of approximately 1.03--1.04~eV across the relevant resonances.

The E1 average combines the eight representatives of the unshifted mesh with their multiplicities $m_r$ and source strengths:
\begin{equation}
 I_{\mathrm{avg}}(E)=\sum_r\frac{m_r}{27}\|b_r\|_S^2 I_r^{\mathrm{unit\ source}}(E),\qquad \sum_r m_r=27.
\end{equation}
The individual spectra retain a common energy scale; no independent peak alignment is applied before averaging. To include the additional multipole response in the two-channel CeO$_2$ comparison, the article uses
\begin{equation}
 \overline I_j(\epsilon)\simeq\langle I_j^{\mathrm{E1}}(\epsilon,\bm k)\rangle_{\bm k}
 +I_j^{\mathrm{E1+E2}}(\epsilon,\Gamma)-I_j^{\mathrm{E1}}(\epsilon,\Gamma).
\end{equation}
Only the E1 term is explicitly averaged over momentum; the remainder at $\Gamma$ is held fixed. Reference-state energy offsets are included afterwards. The white-line separation is 7.1332~eV at $\Gamma$ and 7.30592~eV after averaging. With the calculated sudden weights, the secondary-to-primary white-line peak-height ratio is 1.08799.

\subsection{\texorpdfstring{CePO$_4$}{CePO4} recurrence and momentum tests}
For the primary CePO$_4$ reference, we calculate the dipole response
and the coherent dipole--quadrupole response separately at each of
the 27 points of the shifted $3\times3\times3$ mesh $\mathcal K$.
Each calculation uses 3000 Lanczos iterations. To obtain the mean
spectrum, we first restore each source's transition strength and then
average the intensities at the same energy. For either excitation
channel $a$, this gives
\begin{equation}
 \overline I^{a}(E)=\frac1{27}\sum_{\bm k\in\mathcal K}
 \lVert b_{\bm k}^{a}\rVert_S^2 I_{\rm normalized}^{a}(E,\bm k),
 \qquad a\in\{\mathrm{E1},\mathrm{E1+E2}\}.
 \label{eq:cepo-k-average}
\end{equation}
Here $I_{\rm normalized}^{a}$ is calculated with a unit-norm source;
$\lVert b_{\bm k}^{a}\rVert_S^2$ restores its original strength.
All points use the same screening density and local interactions.
The E1+E2 source contains both excitation amplitudes before propagation;
its response is not obtained by adding separate E1 and E2 intensities.
The individual spectra are neither shifted nor normalized to their own
maxima before averaging, so their relative strengths and k dependence
are retained.

The common energy origin is the maximum of the averaged E1+E2 response
at Lorentzian HWHM 0.3~eV, $-18.57262020$~eV on the internal XSpectruplet
Hamiltonian scale.

Pure E2 is evaluated at $\bm k_1=(1/6,1/6,1/6)$ with 10000 Chebyshev moments over
$[-22,118]$~Ry, Jackson damping and additional Lorentzian broadening.
All detailed panels retain the common white-line zero; no curve is
independently realigned. No full-mesh E2 average is inferred.
The principal E2 maximum is at $-4.13$~eV for HWHM 0.3~eV and
$-4.21$~eV for 0.1~eV on this common white-line-relative scale.

We assess the sensitivity to recurrence length by comparing the first
1000 iterations with the full 3000-iteration result, using the same
broadening and energy origin. The dimensionless peak-normalized RMS
on an energy window $\mathcal W$ is
\begin{equation}
 R_{\mathcal W}=
 \frac{\left[N_{\mathcal W}^{-1}\sum_{E_i\in\mathcal W}
       |y(E_i)-y_{\rm ref}(E_i)|^2\right]^{1/2}}
      {\max_{E_i\in\mathcal W}y_{\rm ref}(E_i)}.
 \label{eq:cepo-rms}
\end{equation}
The uniform grid spacing is 0.01~eV. In Table~\ref{tab:lanczos},
$y_{\rm ref}$ is the L3000 response: the complete mean of
Eq.~\eqref{eq:cepo-k-average} for E1/E1+E2, and $\bm k_1$ for E2.
Each shorter response uses the first 1000 coefficients of the same
recurrence at every contributing point. Intensities and energies are not fitted
independently before comparison.

\begin{table}[H]\centering\small
\caption{Primary CePO$_4$ L1000--L3000 comparison. Entries are
$100R_{\mathcal W}$ over $[-15,20]$~eV, with the L3000 response as
reference. The E1/E1+E2 means use Eq.~\eqref{eq:cepo-k-average};
E2 uses the selected point. The last column gives the absolute change
in its principal maximum.}\label{tab:lanczos}
\begin{tabular}{rrrrr}\toprule
HWHM (eV) & E1 & E2 & E1+E2 & E2 peak change (eV)\\\midrule
0.3 & 0.160 & 0.313 & 0.123 & 0.04\\
0.1 & 1.166 & 2.074 & 0.826 & 0.08\\\bottomrule
\end{tabular}\end{table}

The finite-order convergence of Lanczos and Chebyshev differs
\cite{Mirone2026}. Lanczos rapidly resolves strong localized peaks,
whereas weaker structures and peak positions can vary with recurrence
depth. Chebyshev refines the response globally with increasing order,
and Jackson damping suppresses truncation oscillations at the cost
of finite spectral resolution.

The E2 comparison must distinguish a change of numerical method from a
change of spectral resolution. The usual Lanczos curve has Lorentzian
broadening, whereas the Chebyshev curve also has Jackson smoothing.
To compare the two at the same resolution, we use the finite L3000
tridiagonal matrix to generate a Chebyshev sequence for the same source.
Both this sequence and the directly calculated Chebyshev sequence are
then reconstructed with the same M10000 Jackson kernel and Lorentzian
convolution. Their relative $L^2$ differences are $4.00\times10^{-7}$ and
$5.87\times10^{-7}$ at HWHM 0.3 and 0.1~eV over $[-12,-3]$~eV.
The denominator is the Lanczos-derived profile. No alignment or
independent normalization is applied. This agreement concerns the
specified source, window and resolution, not arbitrarily fine detail.
The usual Lanczos response, without Jackson smoothing,
differs from the Jackson--Chebyshev curve by 2.55\% and 10.27\% in
the same relative norm. These are comparisons of different kernels.

\begin{figure}[H]\centering
\includegraphics[width=0.94\linewidth]{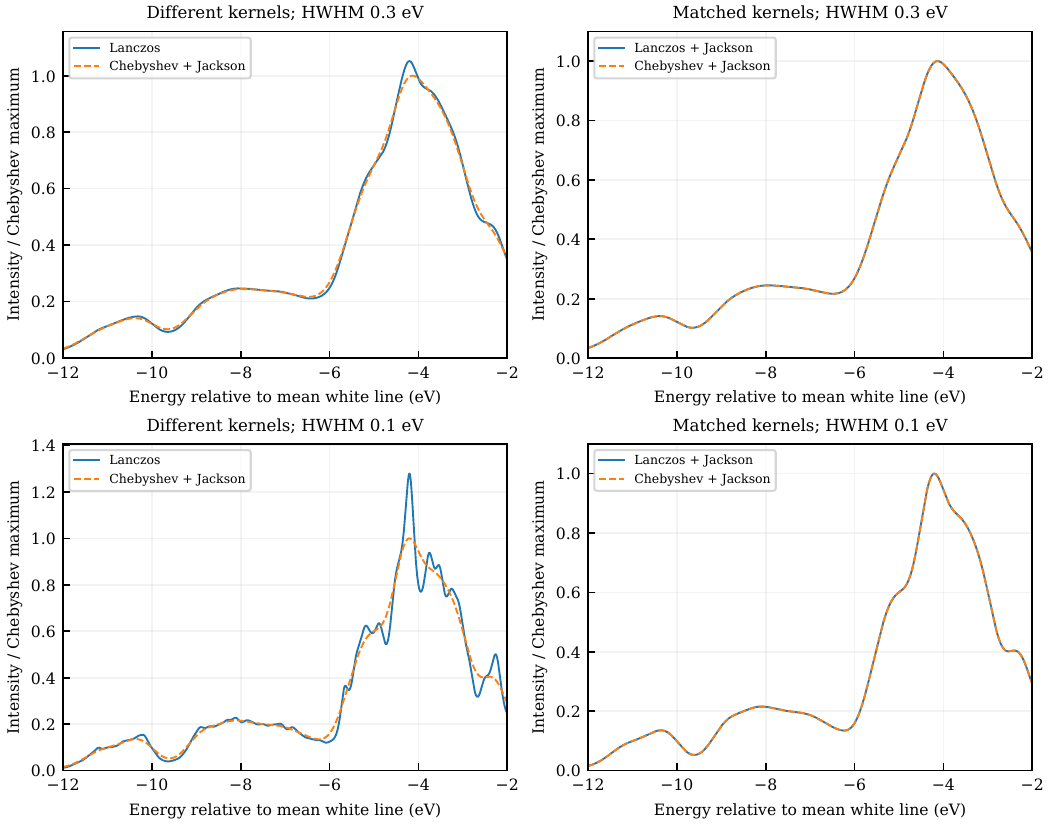}
\caption{Selected-point E2 comparison for the primary CePO$_4$ reference.
Left: L3000 Lanczos and M10000 Jackson--Chebyshev, each
with the indicated Lorentzian HWHM. Right: both sequences reconstructed
with the same M10000 Jackson and Lorentzian kernels; the Lanczos curve
is reconstructed from its finite tridiagonal matrix.
Energies use the mean white line of
Eq.~\eqref{eq:cepo-k-average}, and intensities a common
Chebyshev pre-edge maximum.}\label{fig:methods}
\end{figure}

The E2 sequence satisfies $|\mu_n|\le1.0001\mu_0$ through M10000;
the largest absolute ratio in moments 9001--10000 is 0.0163.
M8000 and M10000 spectra with the same Lorentzian HWHM differ by
0.99\% at 0.3~eV and 2.24\% at 0.1~eV in relative $L^2$ norm over
$[-12,-3]$~eV, using M10000 as reference. Jackson resolution also changes
with expansion order. These checks support the resolved pre-edge
profile but do not remove finite-resolution uncertainty in its
finest structures.

\subsection{Experimental comparison and relative energies}
The experimental curves were digitized from the CeO$_2$ and CePO$_4$ HERFD spectra in Fig.~1 of Ref.~\cite{Kvashnina}. Each is aligned and normalized at its own white-line maximum. The reported combined experimental resolution is approximately 0.8~eV HWHM. These digitized curves provide relative feature positions and a qualitative line-shape comparison; they are not used to fit the screening weights or additional broadening.
The experimental pre-edge--white-line separations are approximately
10.7~eV for CeO$_2$ and 8.1~eV for CePO$_4$. Each compares the two
maxima within the same measured spectrum; the reported precision
reflects their extraction from the published curves. The experimental methods
of Ref.~\cite{Kvashnina} explicitly report 0.8~eV as the half width at
half maximum of the elastic Kapton profile for the combined
incident/emitted-energy resolution.

For the combined CeO$_2$ response, the display cutoff is placed at a local minimum below the $4f$ resonance of each screening background before adding its reference-state offset. The cutoff energies on the respective internal XSpectruplet scales are $-21.6174$~eV for the primary and $-33.7076$~eV for the secondary. The secondary E1-only curve has a dotted continuation to the left axis limit, as discussed in Sec.~\ref{sec:occupied}. The calculated CePO$_4$ response and the experimental curves are shown over their full available data ranges within the plotted axes. Each CePO$_4$ channel retains the same line style throughout its low-energy tail; no occupied-state threshold is imposed by the plotting convention. Numerical comparisons use the stated energy windows without a data mask.
For the weighted comparison, the common plotted energy is
$x_j=E_{{\rm raw},j}+\Delta_j-E_{\rm WL,1}^{\rm avg}$.
Here $E_{{\rm raw},j}$ is the internal XSpectruplet energy for background
$j$, and $\Delta_j$ is its charged QE total cell energy relative to
the primary: $\Delta_1=0$ and $\Delta_2=11.72749209$~eV.
The primary averaged white line has internal energy
$E_{\rm WL,1}^{\rm avg}=-12.34246606$~eV. The two internal cutoffs
therefore become $-9.274939$ and $-9.637684$~eV.
Using the primary $\Gamma$ white line, $-12.22156369$~eV, would instead
give $-9.395841$ and $-9.758586$~eV.
The secondary E1 dotted interval and the solid $4f$ feature are described
in Sec.~\ref{sec:occupied}.

\FloatBarrier
\section{Occupied-state analysis of the pre-edge and lower structures}
\label{sec:occupied}
The connection between an energy-filtered state and the absorption
spectrum follows from their common spectral representation. At fixed
screening background, excitation channel and wave vector, let $H$ be
the self-adjoint spectral Hamiltonian and $|b\rangle$ its excitation
source. If $|\widehat\Psi_\lambda\rangle$ are normalized eigenstates
with energies $\epsilon_\lambda$, a finite-width filter $F_E$ centered
at $E$ gives, before normalization,
\[
 |\Xi_E\rangle=F_E(H)|b\rangle
 =\sum_\lambda F_E(\epsilon_\lambda)
 \langle\widehat\Psi_\lambda|b\rangle|\widehat\Psi_\lambda\rangle.
\]
The same excitation amplitudes enter the source-weighted spectral density,
\[
 I_b(\epsilon)=\sum_\lambda
 |\langle\widehat\Psi_\lambda|b\rangle|^2
 \delta(\epsilon-\epsilon_\lambda),
 \qquad
 \|\Xi_E\|^2=\int d\epsilon\,
 |F_E(\epsilon)|^2 I_b(\epsilon).
\]
The norm in this identity is that of the full coupled state, with its
physical scalar product, rather than the norm of one photoelectron
component. Thus the filter selects the states carrying spectral weight
in the chosen energy interval; its squared norm is an integral weighted
by $|F_E|^2$, not the spectral intensity at a single energy.
For a well-isolated resonance dominated by one eigenstate, normalization
removes the overall excitation amplitude and reveals that state's
character. A group of overlapping levels instead gives a
source-weighted combination, whose interpretation retains the finite
energy resolution. Multiplying the source by an overall constant does
not change the normalized packet or its occupation-weighted overlap.

This analysis asks how much of a selected photoelectron component
resembles KS orbitals already occupied in the screening background.
The weights below are evaluated on the energy-filtered component before
removal of its overlap with occupied KS orbitals. When a packet is
selected for neutral-state construction, that occupied component is
removed and the remainder normalized before injection, as described in
Sec.~\ref{sec:relax}; this subsequent step is distinct from the diagnostic.
At fixed spin $\sigma$ and wave vector $\bm k$, let
$|u_{n\bm k\sigma}\rangle$ be those background KS orbitals, orthonormal
in the PAW metric $S$, and $f_{n\bm k\sigma}$ their occupations.
For the photoelectron component $|\phi_{\alpha_*}\rangle$ in the dominant
local channel $\alpha_*$, the occupation-weighted overlap is
\begin{equation}
 W_{\mathrm{occ}}^{(\alpha_*)}=
 \frac{\sum_n f_{n\bm k\sigma}|\langle u_{n\bm k\sigma}|S|\phi_{\alpha_*}\rangle|^2}
 {\langle\phi_{\alpha_*}|S|\phi_{\alpha_*}\rangle}.
\end{equation}
The squared overlaps, divided by the packet norm, give its fractional
weight in each background KS orbital. With occupations zero or one,
$W_{\rm occ}$ is simply the sum over occupied orbitals: the fraction
of the packet norm lying in the occupied subspace. With fractional
occupations, each orbital's contribution is instead weighted by its
occupation. A partly occupied band therefore contributes an intermediate
amount; this does not classify the whole band as either filled or empty.
We report both conventions for CePO$_4$, rather than treating the
fractional result as the norm of a sharply projected wavefunction.
Each value concerns one local channel of a filtered packet, not the
sum over all local many-electron channels.

The spectral identity above applies to a specified Hamiltonian and source.
The tabulated occupation diagnostics use auxiliary wavepackets constructed
with projector-expanded sources and a projector-based treatment of the
core contribution. The displayed spectra instead use fixed-PAW sources
and direct ionic core action. The calculations share the same screening
backgrounds; corresponding features are identified by excitation channel,
wave vector, internal energy and local orbital character.
This establishes an assignment of corresponding
resonances or energy regions, not an identity of all filtered vectors
between constructions. The tabulated weights characterize the specified
components and do not prescribe a Pauli subtraction from the full
channel-summed spectrum.

\begin{table}[H]\centering\small
\caption{CeO$_2$ occupation weights for packets at $\Gamma$. Energies
refer to the primary $\Gamma$ white line; secondary-state entries include
the charged-reference offset. Unlike Fig.~\ref{main-fig:spectra} of the main article, the
zero here is not the primary averaged white line.}\label{tab:occupied-ceo}
\begin{tabular}{llrr}\toprule
Screening state & Feature & Energy (eV) & $W_{\mathrm{occ}}^{(\alpha_*)}$\\\midrule
Primary & E2 $4f$ pre-edge & $-8.2559$ & $4.4139\times10^{-5}$\\
Primary & Neighboring-Ce $4f$ pre-edge (E1) & $-7.2541$ & $1.1699\times10^{-6}$\\
Primary & White line & 0 & $7.5419\times10^{-6}$\\
Primary & Higher E1 resonance & $+11.4512$ & $2.4351\times10^{-7}$\\
Primary & Lower structure & $-12.7466$ & 0.9999618\\
Primary & Lower structure & $-10.8985$ & 0.9999815\\
Secondary & Lower structure & $-11.9866$ & 0.9999989\\
Secondary & Lower structure & $-14.0765$ & 0.9999990\\
Secondary & E1 maximum A & $-1.5718$ & 0.9998933\\
Secondary & E1 maximum B & $+0.7254$ & 0.9999473\\\bottomrule
\end{tabular}\end{table}
The assigned CeO$_2$ resonances have negligible occupied weight in the components examined, whereas the selected components of the lower structures are almost entirely occupied. This distinction supports the pre-edge assignment and the interpretation of the lower contributions.
The E1 pre-edge at $-7.2541$~eV has $4f$ character on neighboring Ce
sites, distinct from the absorber-centered E2 $4f$ resonance. Its E1
label refers to the dipole-allowed $l=2$ excitation component at the
absorber, not to a local $2p\rightarrow4f$ dipole transition.
This assignment is supported by site- and angular-momentum-resolved
L\"owdin projections onto orthonormalized atomic orbitals. The object
analyzed is the spin-up photoelectron component of the selected E1 packet
at $\Gamma$, after projection onto the unoccupied PAW subspace and
normalization, not a relaxed KS eigenstate or a sum over local channels.
The other Ce sites' $4f$ channels carry 94.98\% of its normalized weight,
while the absorber's $d$ channels carry 1.10\% and its $f$ channels less
than $10^{-8}$. The atomic basis represents 99.57\% of the orbital weight;
the quoted percentages are not renormalized to this represented fraction.
The removed occupied weight is $1.17\times10^{-6}$.
These projections concern the diagnostic packet described above,
not the source-resolved components of the displayed spectrum.

The two additional secondary E1 maxima were checked with 10000-moment filters at internal energies $-25.52082$ and $-23.22368$~eV, or $-8.70497$ and $-6.40783$~eV relative to the secondary $\Gamma$ E1 white line. Their dominant components have spin down and spin up, respectively. Projection onto the 384 occupied KS orbitals of the matching spin leaves Pauli-allowed weights $1.06718\times10^{-4}$ and $5.26951\times10^{-5}$. Their PAW norms differ from unity by less than $10^{-14}$ and the occupied-space orthonormality residual is below $2\times10^{-13}$.
These filter centers coincide, at the resolution of the energy grid,
with the corresponding local maxima of the secondary $\Gamma$ E1
response used for the spectra. This supplies the energy-and-channel
correspondence for the two assignments.

These two E1 features, whose dominant diagnostic components have nearly
unit occupied weights, lie above the secondary local quadrupolar $4f$
resonance. In the main figures their region is shown as dotted segments between
neighboring minima. The E1 continuation is dotted to the left axis limit;
in the combined response it stops at the minimum following the
lower-energy $4f$ resonance, which remains solid. The E2 assignment is
based on its local quadrupolar character and the nearly empty local
$4f$ subspace; no direct $W_{\rm occ}$ evaluation for this secondary
E2 packet is available. The dotted interval is a display convention,
not a channel-summed Pauli projection.

For the primary CePO$_4$ reference, the analysis uses 8000-moment E2
packets at $\bm k=(1/6,1/6,1/6)$. The table compares the actual
Gaussian occupations with a projection onto the first 82 spin-up orbitals
occupied with unit weight. All three selected components have spin up.
\begin{table}[H]\centering\small
\caption{CePO$_4$ conditional occupied weights under two occupation
conventions. Energies are relative to the primary $27$-point-averaged
E1+E2 white-line maximum at HWHM 0.3~eV in the diagnostic calculation:
$-18.31262020$~eV on the internal XSpectruplet scale, before any estimated
relaxation shift. This zero is 0.26~eV above the white-line zero used
for the spectra in Fig.~\ref{main-fig:paired-materials} of the main article.}\label{tab:occupied-cepo}
\begin{tabular}{rrr}\toprule
Energy (eV) & Fractional occupations & First 82 orbitals, unit occupation\\\midrule
$-10.0$ & 0.711533 & 0.667813\\
$-6.0$ & 0.258732 & 0.105610\\
$-4.2$ & 0.201107 & 0.087621\\\bottomrule
\end{tabular}\end{table}
On the white-line-relative axis of the main-figure spectra, the same
three fixed internal filter energies are $-9.74$, $-5.74$ and
$-3.94$~eV. They sample the lower-energy wing and the principal pre-edge
region, whose displayed E2 maximum is at $-4.13$~eV for HWHM 0.3~eV.
They characterize these energy regions rather than assigning an
individual eigenstate to every resolved multiplet feature.
The 200 saved KS states capture at least 99.9928\% of each analyzed
amplitude. The PAW Gram error is below $1.1\times10^{-13}$ and the
reconstructed beta projections agree with QE below $8\times10^{-15}$
relative error. At $-4.2$~eV, bands 84 and 83 contribute 42.98\% and
27.81\% of the dominant-component norm, with occupations 0.1423 and
0.1976. At $-6$~eV, band 83 contributes 75.91\% of the norm. Thus this
component is mostly unoccupied; the component at $-10$~eV is mixed.
These conditional weights do not determine a channel-summed blocked
intensity or a sharp occupied-state boundary. The full calculated
CePO$_4$ response is retained, with no mask or line-style boundary
in its low-energy tail.

\FloatBarrier
\section{Local multiplet Hamiltonian and operator conventions}
\label{sec:operators}
This section documents corrections to the XSpectruplet implementation
introduced in the present work, together with the operator conventions
needed to specify them. The corrections concern the repeated QE
contribution of equivalent open-shell spectators, the placement of local
one-electron and double-counting terms in the different many-electron
spaces, and the action of core spin--orbit coupling in the coupled PAW
representation.

For the open-shell calculation, the local electron--electron interaction and the mean one-electron contribution already included in the QE background must be counted consistently. Two distinct corrections are used: subtraction of the nonspherical Hartree--Fock field associated with the explicit local interaction, and removal of the repeated mean QE energy associated with equivalent spectator placements.

For $n$ equivalent localized spectators, the mean shell energy is evaluated as
\begin{equation}
 \overline\epsilon_{\mathrm{open}}^{\mathrm{QE}}
 =\frac{1}{N_{\mathrm{spin}}N_{\mathrm{open}}}
 \sum_{\sigma\eta}
 \frac{\langle\widetilde\phi_{\eta\sigma}|H_{\mathrm{QE}}^\sigma|\widetilde\phi_{\eta\sigma}\rangle}
 {\langle\widetilde\phi_{\eta\sigma}|S|\widetilde\phi_{\eta\sigma}\rangle}.
\end{equation}
In the equivalent-placement representation, the counterterm has coefficient
\begin{equation}
 -\overline\epsilon_{\mathrm{open}}^{\mathrm{QE}}\frac{\sqrt{n(n+1)}}{n+1}.
\end{equation}
It vanishes for the $4f^0$ local sector in CeO$_2$ and is active for the $4f^1$ spectator in CePO$_4$.
Let $|\widetilde\phi_\eta\rangle$ be the first-shell PAW partial waves,
$d_\eta=\langle\widetilde\phi_\eta|S|\widetilde\phi_\eta\rangle$, and define
the normalized atomic projection used here by
\begin{equation}
 P=\sum_\eta
 \frac{|\widetilde\phi_\eta\rangle
       \langle\widetilde\phi_\eta|S}{d_\eta}.
\end{equation}
The channel-dependent Pauli exclusions are applied before this projection.
For a channel with $n$ additional equivalent spectators, its stored atomic
component is $v_{\rm at}=\sqrt{n/(n+1)}\,P\psi$. The counterterm therefore
acts, before summing equivalent placements, as
\begin{equation}
 C_n\psi=-\overline\epsilon_{\rm open}^{\rm QE}
 \sqrt{\frac{n}{n+1}}\,v_{\rm at}
 =-\overline\epsilon_{\rm open}^{\rm QE}\frac{n}{n+1}P\psi.
\end{equation}
If $\mathcal R=I+\sum_{a=1}^{n}\mathcal E_a$ is the redistribution map,
with the fermionic signs included in the equivalent-placement maps
$\mathcal E_a$, the correction in the propagated equation is
$\mathcal R C_n$. On the equivalent local component,
$\mathcal R P\psi=(n+1)P\psi$, giving
$-n\overline\epsilon_{\rm open}^{\rm QE}P\psi$.
It removes the $n$ spectator repetitions while retaining the
photoelectron's own mean QE term. Channels with $n=0$ are unchanged.

The local core-spin--orbit interaction is part of the excited-state
Hamiltonian,
\begin{equation}
 H_{\mathrm{loc}}^{\mathrm{exc}}=
 H_{\mathrm{loc}}^{\mathrm{rest}}+H_{\mathrm{SO}}^{\mathrm{core}}.
\end{equation}
It acts on the local core-hole degrees of freedom and as the identity
on the photoelectron coordinate. In the channel representation this reads
\begin{equation}
 (\mathcal K_{\rm core}\Psi)_\alpha(\bm r)
 =\sum_\beta (K_{\rm core})_{\alpha\beta}\psi_\beta(\bm r).
\end{equation}
Here $(K_{\rm core})_{\alpha\beta}$ are the core-spin--orbit matrix
elements between local channels. The coefficient of this interaction
is varied in the sensitivity tests of Sec.~\ref{sec:source}; it does
not change the underlying self-consistent screening density.

For CePO$_4$, an empirical local field selects the initial $4f^1$ state. The nonzero $f$-channel field amplitudes are 0.04 and 0.02~eV, with a small orientation field $B_z=-10^{-6}$~eV. The spin-averaged orbital density is
\begin{equation}
 \overline M_{ij}=\tfrac12\sum_\sigma\langle g|f_{i\sigma}^{\dagger}f_{j\sigma}|g\rangle,
 \qquad \operatorname{Tr}\overline M=\tfrac12.
\end{equation}
The same orbital matrix is used in the two spin blocks of the local correction, giving one spectator electron in total. The nonspherical field subtraction is included in the excited local Hamiltonian without redefining the initial empirical-field eigenstate from which the excitation source is constructed. The Slater-integral reduction is the same as for the explicit interaction.

With the convention $M_{ij}=\langle f_i^{\dagger}f_j\rangle$ and $f_{\mathrm{sph}}=Uf_{\mathrm{real}}$, the density transforms as
\begin{equation}
 M_{\mathrm{sph}}=U^*M_{\mathrm{real}}U^T.
\end{equation}
The corresponding one-electron contraction is $\sum_{ij}h_{ij}M_{ij}$. This density convention is stated explicitly to distinguish it from the convention in which the two density indices are reversed.

The empirical field is a sum over the nine Ce--O bonds. For each bond
$\bm b$, rotate the $l=3$ basis so that its local $z$ axis points along
$\widehat{\bm b}$. With normalized spherical harmonics in the
Condon--Shortley convention and rotation matrix $D^{(3)}(R_{\bm b})$,
\begin{equation}
 H_{\rm CF}=\sum_{\bm b}
 \left(\frac{|\bm b|}{d_{\rm ref}}\right)^{-3}
 D^{(3)}(R_{\bm b})
 \left[V_0|0\rangle\langle0|+
 V_1\!\!\sum_{m=\pm1}|m\rangle\langle m|\right]
 D^{(3)}(R_{\bm b})^\dagger .
\end{equation}
Here $V_0=0.04$~eV and $V_1=0.02$~eV; the $|m|=2,3$ amplitudes are zero.
The reference distance is the root-mean-square bond length,
$d_{\rm ref}=(\sum_{\bm b}|\bm b|^2/9)^{1/2}=2.55745$~\AA.
The bonds are the nine nearest periodic oxygen neighbors of the absorber
in the Cartesian cell of Sec.~\ref{sec:settings}, with lengths
2.4454--2.7828~\AA. The matrix is transformed to the real QE harmonic
basis before constructing the local many-electron Hamiltonian.
In spherical order $m=-3,\ldots,3$, the real basis is ordered as
$|0\rangle,|1c\rangle,|1s\rangle,\ldots,|3c\rangle,|3s\rangle$, with
\[
 |mc\rangle=\frac{|-m\rangle+(-1)^m|m\rangle}{\sqrt2},\qquad
 |ms\rangle=\frac{i[|-m\rangle-(-1)^m|m\rangle]}{\sqrt2}.
\]
This bond-field construction follows the empirical geometry-dependent
approach of Ref.~\cite{Longo2022}.

For completeness, define the spin-orbital Coulomb tensor by
$\widehat V=\sum_{abcd}V_{abcd}f_a^\dagger f_b^\dagger f_c f_d$.
The intrashell tensor includes the factor $1/2$ and the reduced
$F^2,F^4,F^6$ integrals. With $M_{ij}=\langle f_i^\dagger f_j\rangle$,
its Hartree--Fock functional and derivative are
\begin{align}
 E_{\rm HF}[M]&=\sum_{abcd}V_{abcd}
       (M_{ad}M_{bc}-M_{ac}M_{bd}),\\
 \Sigma_{ij}[M]&=2\sum_{kl}
       (V_{iklj}-V_{ikjl})M_{kl}.
\end{align}
The factor two follows from differentiating the quadratic functional;
it is not an additional spin degeneracy. The derivative was checked by
symmetric finite differences on complex Hermitian density variations.
The same 0.75 Slater reduction is used for $V$ and for the explicit
local interaction. The monopole $F^0$ term is not subtracted through
this nonspherical field.

The correction is $C[M]=-\Sigma[M]-a_c I$, where the filled-core exchange
average retained by this convention is
\[
 a_c=-(2l_c+1)\sum_K G^K(c,f)
 \begin{pmatrix}l_c&K&3\\0&0&0\end{pmatrix}^{\!2}.
\]
The sum follows the allowed angular couplings and uses the same Slater
reduction. The scalar energy convention is fixed when exporting the
local Hamiltonians: the mean diagonal energy is subtracted separately
from each photoelectron-shell block of the fully local core-excited
Hamiltonian, and the trace divided by the dimension is subtracted from
the residual-ion Hamiltonian. These operations define the local
multiplet energy splittings; they do not subtract the mean of the QE
one-electron potential.

For the two CeO$_2$ screening channels, the same exported local matrices
and core-spin--orbit parameters are used, so these local energy
conventions are common to both. The QE contribution retains its periodic
potential convention, with Hartree component $V_{\rm H}(\bm G=0)=0$
and the ionic contribution fixed by the identical PAW pseudopotentials
and cell. The two references also have the same electron count and net
charge. Neither Hamiltonian is shifted to its own Fermi level or
white-line maximum. The internal spectral energies can thus be combined
with the charged-reference total-energy differences as in
Eq.~\eqref{main-eq:white-line-accounting} of the main article, followed
by one overall shift common to both channels. This fixes the numerical
energy convention within the frozen-background, independent-channel
description; it is not a calculation of the absolute absorption energy.

In the CePO$_4$ calculation, the empirical crystal field acts
only in the initial local $4f^1$ Hamiltonian that determines $|g\rangle$.
The correction $C[M]$ acts in the fully local core-excited Hamiltonian
and in the residual-ion Hamiltonian coupled to the plane-wave
photoelectron; it does not modify this initial empirical-field eigenstate.
For E2, these two core-excited spaces are $2p^54f^2$ and $2p^54f^1$,
respectively. For E1, they are $2p^54f^15d^1$ and $2p^54f^1$.

An optional construction, not used for this CePO$_4$ reference, supplies
a physical Wannier Hamiltonian $h_W$. Its corrected one-electron term
$h_W+C[M]$ then enters the initial local Hamiltonian, whereas the
fully local core-excited and residual-ion Hamiltonians receive only
$C[M]$. The physical Wannier and empirical crystal fields are thus
not added a second time to the local terms used during propagation.

\FloatBarrier
\section{\texorpdfstring{CePO$_4$}{CePO4}: transferred gap estimate and a separate multiplet check}
\label{sec:centroid}

\subsection{Transferred gap estimate}
The adopted estimate shifts the CePO$_4$ $4f$ pre-edge down by
2.250~eV relative to the white line, without changing the internal
multiplet structure. This shift comes from the CeO$_2$ comparison in
Sec.~\ref{sec:relax}: its fixed-background spectral gap is 8.255905~eV
at $\Gamma$, whereas the independently relaxed neutral $4f$ and
white-line states differ in total energy by approximately 10.506~eV.
The estimated correction is the difference between these two gaps:
\begin{equation}
 \Delta_{\mathrm{est}}
 \simeq(10.506-8.255905)~\mathrm{eV}
 \simeq2.250~\mathrm{eV}.
\end{equation}
Assuming a comparable net gap correction in CePO$_4$ gives
\begin{equation}
 G_{\mathrm{CePO_4}}^{\mathrm{est}}
 \simeq G_{\mathrm{CePO_4}}^{\mathrm{spec}}+\Delta_{\mathrm{est}}
 \simeq(4.13+2.25)~\mathrm{eV}=6.38~\mathrm{eV}.
\end{equation}
Here $G$ is the positive white-line--pre-edge separation. For CePO$_4$,
the spectral value uses the selected-point E2 maximum and the primary
averaged E1+E2 white line of Eq.~\eqref{eq:cepo-k-average}.
Both have Lorentzian HWHM 0.3~eV, with the additional M10000 Jackson
kernel retained in the E2 curve.
The estimated peak is therefore at $-6.38$~eV on that calculated
white-line-relative axis. Experiment gives approximately $-8.1$~eV
relative to its own measured white line, about 1.7~eV lower.

This transfers the estimated net correction to a relative peak separation;
no absolute energy is transferred between materials. Both optimized CeO$_2$
endpoints enter its definition. The transfer to CePO$_4$ is heuristic and
does not follow from a pair of relaxed neutral states for its current
reference. Preserving the multiplet structure under this shift is an
additional approximation.

\subsection{Local-multiplet shape check}
\label{sec:local-shape}
The local reference retains only the $2p$ core and the first $4f$
radial shell, with seven active angular projectors. Auxiliary $5f$,
$5d$ and $6d$ shells are excluded from both the active Hamiltonian
and the excitation source, not merely from the initial wavepacket.
The PW Hamiltonian and continuum contributions are omitted.
The local Coulomb, spin--orbit and density-derived one-body parameters
are the same as in the full calculation.

An independent diagonalization of the antisymmetric $2p^54f^2$
Hamiltonian, of dimension 546, provides a source-weighted atomic
reference. The PW implementation restricted to this local space,
evaluated with 1000 Lanczos steps, agrees with that reference to
relative $L^2$ differences of $1.72\times10^{-6}$ and
$4.69\times10^{-6}$ at Lorentzian HWHM 0.3 and 0.1~eV.
Both profiles have unit integrated $L_3$ weight; no independent energy
alignment is used in this implementation check. Their $L_3$ centroids
differ by $3.15\times10^{-7}$~eV.

For the local tridiagonal matrix $T$, diagonalization
$TV_\nu=\epsilon_\nu V_\nu$ gives strengths
$w_\nu=|(V_\nu)_1|^2$, with $\sum_\nu w_\nu=1$.
The 1000 quadrature poles, including negligible or repeated entries,
are not 1000 distinct physical multiplet states. We define
\begin{equation}
 \overline E_{L_3}^{\rm loc}
 =\frac{\sum_{\epsilon_\nu<E_{\rm sep}}w_\nu\epsilon_\nu}
 {\sum_{\epsilon_\nu<E_{\rm sep}}w_\nu}.
\end{equation}
The resolved core gap permits $E_{\rm sep}=17.9004$~eV on the local
Hamiltonian's internal scale; the weight within 3~eV of this separator
is below $10^{-6}$. The resulting $L_3$ weight is 0.714419445 and
its centroid is $-32.96685050$~eV on that same internal scale, not
on the photon-energy or white-line-relative scale. At HWHM 0.3~eV,
\begin{equation}
 \delta_{\rm mult}
 =E_{\max}^{\rm loc}-\overline E_{L_3}^{\rm loc}
 \simeq-0.610~{\rm eV}.
\end{equation}

For the local--continuum shape comparison, both curves use a common
Lorentzian HWHM of 0.3~eV. The continuum profile is reconstructed
from the verified selected-point L3000 calculation, without the
additional Jackson smoothing used for the displayed E2 spectrum.
Its maximum is $-4.20$~eV relative to the averaged white line
defined in Sec.~S6. We align this maximum with the local maximum:
\begin{equation}
 a=E_{\max}^{\rm spec}-\delta_{\rm mult}
   =-4.20-(-0.610)=-3.590~{\rm eV}.
\end{equation}
The continuum energy is displayed as $x_{\rm aligned}=x_{\rm WL}-a$,
while the local poles are referred to their own centroid.
The parameter $a$ is a shape-alignment offset, not a measured continuum
centroid or a relaxation energy; it does not enter the gap estimate.

The curves are sampled at 0.0025-eV spacing and normalized to unit area
over $[-12,2]$~eV on the aligned axis. Their relative $L^2$ difference,
with the local profile as reference, is
\begin{equation}
 \delta_{\rm shape}=
 \frac{\left[\sum_i|y_{\rm full}(E_i)-y_{\rm loc}(E_i)|^2\right]^{1/2}}
      {\left[\sum_i|y_{\rm loc}(E_i)|^2\right]^{1/2}}.
\end{equation}
The common grid spacing cancels. The difference is 47.8\%, with
Pearson correlation 0.8925994. The local multiplet captures a correlated
part of the pre-edge profile, while the full response also contains
weight associated with its electronic environment and continuum.
It can also contain occupied-background components, since no global
empty-state projection is applied, as discussed in Sec.~\ref{sec:occupied}.
This is a shape diagnostic after explicit peak alignment, not an
uncertainty in the gap or a derivation of the heuristic relaxation shift.

\FloatBarrier